\documentclass{llncs}

\usepackage{xspace}
\usepackage{graphicx}
\usepackage{todonotes}
\usepackage{bbm}
\usepackage{paralist}
\usepackage{amsmath}
\usepackage{amsfonts,amssymb,latexsym}

\newcommand{\N}{\mbox{I$\!$N}}

\newcommand{\zug}[1]{\langle #1  \rangle}
\newcommand{\tup}[1]{\langle #1 \rangle}

\newcommand{\stam}[1]{}

\newcommand{\A}{{\cal A}}
\renewcommand{\S}{{\cal S}}
\newcommand{\B}{{\cal B}}

\newcommand{\D}{{\cal D}}

\newcommand{\T}{{\cal T}}

\newcommand{\RR}{\mathbb{R}}
\newcommand{\EE}{\mathbb{E}}
\newcommand{\remove}[1]{}

\newcommand{\psis}{\psi_{\it strong}}
\newcommand{\psiw}{\psi_{\it weak}}

\newcommand{\sem}[1]{[\![#1]\!]}
\newcommand{\True}{\mathtt{True}}
\newcommand{\False}{\mathtt{False}}
\newcommand{\LTL}{{\ensuremath{\rm LTL}}\xspace}
\newcommand{\EQLTL}{{\ensuremath{\rm EQLTL}}\xspace}
\newcommand{\AQLTL}{{\ensuremath{\rm AQLTL}}\xspace}

\newcommand{\comb}{{\mathtt{comb}}}

\newcommand{\condvar}{{\rm cond}}

\newcommand{\Next}{\mathsf{X}}

\newcommand{\Ev}{\mathsf{F}}
\newcommand{\Alw}{\mathsf{G}}
\newcommand{\Until}{\mathsf{U}}

\newcommand{\avg}[1]{\oplus_{#1}}
\newcommand{\Fu}[2]{\ensuremath {{#1}{[#2]}}\xspace}

\newcommand{\F}{{\cal F}}
\newcommand{\FLTL}{\Fu{\LTL}{\F}\xspace}
\newcommand{\factorU}{\triangledown} 
\newcommand{\EQFLTL}{\Fu{\EQLTL}{\F}}
\newcommand{\AQFLTL}{\Fu{\AQLTL}{\F}}

\newcommand{\IOo}{(2^{I\cup O})^\omega}
\newcommand{\Io}{(2^{I})^\omega}

\newcommand{\ges}{{\sc ge}\xspace}

\title{Good-Enough Synthesis}
\author{Shaull Almagor\inst{1}
	\thanks{Supported  by the European Union's Horizon 2020 research and innovation programme under the Marie Sk{\l}odowska-Curie grant agreement No. 837327.} 
	\and Orna Kupferman\inst{2}
	\thanks{Supported in part by the Israel Science Foundation, grant No. 2357/19.}
}
\institute{Department of Computer Science, Technion, Israel.
	\and
	School of Computer Science and Engineering, The Hebrew University, Israel.
}
\begin{document}
	
	\pagestyle{plain}
	\pagenumbering{arabic}
	\maketitle
	\begin{abstract}
		In the classical {\em synthesis\/} problem, we are given an LTL formula $\psi$ over sets of input and output signals, and we synthesize a system $\T$ that {\em realizes}~$\psi$: with every input sequences $x$, the system associates an output sequence $\T(x)$ such that the generated computation $x \otimes \T(x)$ satisfies $\psi$. In practice, the requirement to satisfy the specification in all environments is often too strong, and it is common to add assumptions on the environment.
		We introduce and study a new type of relaxation on this requirement. In \emph{good-enough synthesis} (\ges-synthesis), the system is required to generate a satisfying computation only if one exists. Formally, an input sequence $x$ is \emph{hopeful} if there exists some output sequence $y$ such that the computation $x \otimes y$ satisfies $\psi$, and a system \ges-realizes $\psi$ if it generates a computation that satisfies $\psi$ on all hopeful input sequences. \ges-synthesis is particularly relevant when the notion of correctness is {\em multi-valued} (rather than Boolean), and thus we seek systems of the highest possible quality, and when synthesizing {\em autonomous systems}, which interact with unexpected environments and are often only expected to do their best. 
		
		We study \ges-synthesis in Boolean and multi-valued settings. In both, we suggest and solve various definitions of \ges-synthesis, corresponding to different ways a designer may want to take hopefulness into account.  We show that in all variants, \ges-synthesis is not computationally harder than traditional synthesis, and can be implemented on top of existing tools. Our algorithms are based on careful combinations of nondeterministic and universal automata. We augment systems that \ges-realize their specifications by monitors that provide satisfaction information. In the  multi-valued setting, 
		we provide both a worst-case analysis and an expectation-based one, the latter corresponding to an interaction with a stochastic environment.
	\end{abstract}

	\section{Introduction}
	\label{intro}
	{\em Synthesis\/} is the automated construction of a system from its specification: given a specification $\psi$, typically by 
	%short
	a
	%means of a 
	linear temporal logic (LTL) formula  over sets $I$ and $O$ of input and output signals, the goal is to construct a finite-state system that satisfies $\psi$ \cite{Chu63,PR89a}. 
	At each moment in time, the system reads an assignment, generated by the environment, to the signals in $I$, and responds with an assignment to the signals in $O$. Thus, with every input sequence, the system associates an output sequence. 
	The system {\em realizes\/} $\psi$ if $\psi$ is satisfied in all the interactions of the system, with all environments~\cite{BCJ18}. 
	
	In practice, the requirement to satisfy the specification in all environments is often too strong. Accordingly, it is common to add assumptions on the behavior of the environment. An assumption may be direct, say given by an LTL formula that restricts the set of possible input sequences \cite{CHJ08}, less direct, say a bound on the size of the environment \cite{KLVY11} or other resources it uses, or conceptual, say rationality from the side of the environment, which may have its own objectives \cite{FKL10,KPV16}. We introduce and study a new type of relaxation of the requirement to satisfy the specification in all environments. The idea behind the relaxation is that if an environment is such that no system can interact with it in a way that satisfies the specification, then we cannot expect our system to succeed. In other words, the system has to satisfy the specification only when it interacts with environments in which this mission is possible. This is particularly relevant when synthesizing {\em autonomous systems}, which interact with unexpected environments and often replace human behavior, which is only expected to be {\em good enough} \cite{Win71}, and when the notion of correctness is multi-valued (rather than Boolean), and thus we seek {\em high-quality\/} systems. 
	
	Before we explain the relaxation formally, let us consider a simple example, and we start with the Boolean setting. 
	Let $I=\{{\it req}\}$ and $O=\{{\it grant}\}$. Thus, the system receives requests and generates grants. Consider the specification  
	$\psi= \Alw \Ev({\it req} \wedge {\it grant}) \wedge \Alw \Ev (\neg {\it req} \wedge \neg {\it grant})$. Clearly, $\psi$ is not realizable, as an input sequence need not satisfy $\Alw \Ev{\it req}$ or $\Alw\Ev \neg {\it req}$. However, a system that always generates a grant upon (and only upon) a request, {\em \ges-realizes\/} $\psi$, in the sense that for every input sequence, if there is some interaction with it with which $\psi$ is satisfied, then our system generates such an interaction. 
	
	Formally, we model a system by a strategy $f:(2^I)^+ \rightarrow 2^O$, which given an input sequence $x=i_0 \cdot i_1 \cdot i_2 \cdots \in (2^I)^\omega$, generates an output sequence $f(x)=f(i_0) \cdot f(i_0 \cdot i_1) \cdot f(i_0 \cdot i_1 \cdot i_2) \cdots \in (2^O)^\omega$, inducing the computation $x \otimes f(x)=(i_0\cup f(i_0)) \cdot (i_i \cup f(i_0\cdot i_1)) \cdot (i_2\cup f(i_0 \cdot i_1\cdot i_2)) \cdots \in (2^{I \cup O})^\omega$, obtained by ``merging" $x$ and $f(x)$. In traditional realizability, a system realizes $\psi$ if $\psi$ is satisfied in all environments. Formally, for all input sequences $x \in (2^I)^\omega$, the computation $x \otimes f(x)$ satisfies $\psi$. For our new notion, we first define when an input sequence $x \in (2^I)^\omega$ is {\em hopeful}, namely there is an output sequence $y \in (2^O)^\omega$ such that the computation $x \otimes y$ satisfies $\psi$. Then, a system {\em \ges-realizes\/} $\psi$ if $\psi$ is satisfied in all interactions with hopeful input sequences. Formally, for all  $x\in (2^I)^\omega$, if $x$ is hopeful, then the computation $x \otimes f(x)$ satisfies $\psi$.
	
	Since LTL is Boolean, synthesized systems are correct, but there is no reference to their quality. This is a crucial drawback, as designers would be willing to give up manual design only if automated-synthesis algorithms return systems of comparable quality. Addressing this challenge, researchers have developed quantitative specification formalisms. For example, in \cite{BCHJ09}, the input to the synthesis problem includes also Mealy machines that grade different realizing systems. In~\cite{ABK16}, the specification formalism is the multi-valued logic \FLTL, which augments \LTL with quality operators. The satisfaction value of an \FLTL formula is a real value in $[0,1]$, where the higher the value, the higher the quality in which the computation satisfies the specification.  The quality operators in~$\F$ can prioritize and weight different scenarios. 
	%short
	The synthesis algorithm for \FLTL seeks systems with a highest possible satisfaction value. 
	%Using a multi-valued specification formalism, synthesis is upgraded to generate not only correct, but also high-quality systems. In particular, the synthesis algorithm for \FLTL seeks systems with a highest possible satisfaction value. 
	One can consider either a worst-case approach, where the satisfaction value of a system is the satisfaction value of its computation with the lowest satisfaction value \cite{ABK16}, or a stochastic approach, where it is the expected satisfaction value, given a distribution of the inputs \cite{AK16}. 
	%Since its introduction, \FLTL has been extended with {\em discounting operators} \cite{ABK16,NH15}, and was used  in order to reason about quality and fuzziness of strategic behaviours.\cite{BKMMMP19}.

	%\begin{example}
	%	\label{xmp: intro autonomous car}
	We demonstrate the effectiveness of \ges-synthesis in the multi-valued setting with an example. 
	Consider an acceleration controller of an autonomous car. Normally, the car should maintain a relatively constant speed. However, in order to optimize travel time, if a long stretch of road is visible and is identified as low-risk, the car should accelerate. Conversely, if an obstacle or some risk factor is identified, the car should decelerate. %Both acceleration and deceleration should ideally be gradual, but it is acceptable (to a lesser degree) that they are rapid. Finally, clearly the car cannot accelerate and decelerate at the same time.
	Clearly, the car cannot accelerate and decelerate at the same time.	
	We capture this desired behavior with the following \FLTL formula over the inputs $\{\textit{safe},\textit{obs}\}$ and outputs $\{\textit{acc},\textit{dec}\}$:
	\[\psi=\Alw (\textit{safe}\to (\textit{acc}\avg{\frac23}\Next\textit{acc}))
	\wedge 
	\Alw (\textit{obs}\to (\textit{dec}\avg{\frac34}\Next\textit{dec}))
	\wedge
	\Alw (\neg (\textit{acc}\wedge \textit{dec})).
	\]
	Thus, in order to get satisfaction value $1$, each detection of a safe stretch should be followed by an acceleration during two transactions, with a preference to the first (by the semantics of the weighted average $\avg{\lambda}$ operator, the satisfaction value of $\textit{safe}\to (\textit{acc}\avg{\frac23}\Next\textit{acc})$ is $1$ when $\textit{safe}$ is followed by two $\textit{acc}$s, $\frac{2}{3}$ when it is followed by one $\textit{acc}$, and $\frac{1}{3}$ if it is followed by one $\textit{acc}$ with a delay), and each detection of an obstacle should be followed by a deceleration during two transactions, with a (higher) preference to the first. Clearly, $\psi$ is not realizable with satisfaction value $1$, as for some input sequences, namely those with simultaneous or successive occurrences of \textit{safe} and \textit{obs}, it is impossible to respond with the desired patterns of acceleration or declaration. Existing frameworks for synthesis cannot handle this challenge. Indeed, we do not want to add an assumption about \textit{safe} and \textit{obs} occurring far apart. Rather, we want our autonomous car to behave 
	%in an optimal way 
	as well as possible
	also in problematic environments, and we want, when we evaluate the quality of a car, to take into an account the challenge posed by the environment. This is exactly what high-quality \ges-synthesis does: for each input sequence, it requires the synthesized car to obtain the maximal satisfaction value that is possible for that input sequence. 	
	
	We show that in the Boolean setting, \ges-synthesis can be reduced to synthesis of LTL with quantification of atomic 
	propositions \cite{SVW87}. Essentially, \ges-synthesis of $\psi$ amounts to synthesis of $(\exists O. \psi) \rightarrow \psi$.  We show that by carefully switching between nondeterminisitc and universal automata, we can solve the \ges-synthesis problem in doubly-exponential time, thus it is not harder than traditional synthesis. Also, our algorithm is {\em Safraless}, thus no determinization and parity games are needed \cite{KV05c,KPV06}. 
	
	A drawback of \ges-synthesis is that we do not actually know whether the specification is satisfied. We describe two ways to address this drawback. 
	The first goes beyond providing satisfaction information and enables the designer to partition the specification into a {\em strong\/} component, which is guaranteed to be satisfied in all environments, and a {\em weak\/} component, which is guaranteed to be satisfied only in hopeful ones. The second way augments \ges-realizing systems by ``satisfaction indicators". For example, we show that when a system is lucky to interact with an environment that generates a prefix of an input sequence such that, when combined with a suitable prefix of an output sequence, the specification becomes realizable, then \ges-synthesis guarantees that the system indeed  responds with a suitable prefix of an output sequence. Moreover, it is easy to add to the system a monitor that detects such prefixes, thus indicating that  the specification is going to be satisfied in all environments. While a naive construction of such a monitor is based on a solution of the synthesis problem, we show that since the system \ges-realizes the specification, the monitor can be based on a solution of the {\em universal-satisfiability} problem (that is, deciding whether all input sequences are hopeful), which is much simpler. Additional monitors we suggest detect prefixes after which the satisfaction becomes valid or unsatisfiable. 
	
	We continue to the quantitative setting. We parameterize hope by a satisfaction value $v \in [0,1]$ and say that an input sequence $x\in \Io$ is {\em $v$-hopeful\/} for an \FLTL formula $\psi$ if an interaction with it can generate a computation that satisfies $\psi$ with value at least $v$. Formally, there is an output sequence $y \in (2^O)^\omega$ such that $\sem{x \otimes y,\psi} \geq v$, where for a computation $w \in (2^{I \cup O})^\omega$, we use $\sem{w,\psi}$ to denotes the satisfaction value of $\psi$ in $w$. As we elaborate below, while the basic idea of \ges-synthesis, namely ``input sequences with a potential to high quality should realize this potential" is as in the Boolean setting, there are several ways to implement this idea. 
	
	We start with a worst-case approach. There, a strategy $f:(2^I)^+ \rightarrow 2^O$ \ges-realizes an \FLTL formula $\psi$ if for all input sequences $x \in (2^I)^\omega$, if $x$ is $v$-hopeful, then $\sem{x \otimes f(x),\psi} \geq v$. The requirement can be applied to a threshold value or to all values $v \in [0,1]$. 
	For example, our autonomous car controller has to achieve satisfaction value $1$ in roads with no simultaneous or successive occurrences of \textit{safe} and \textit{obs}, and value $\frac{3}{4}$ in roads that violate the latter only with some \textit{obs} followed by \textit{safe}.
	We then argue that the situation is similar to that of {\em high-quality assume guarantee synthesis\/} \cite{AKRV17}, where richer relations between a quantitative assumption and a quantitative guarantee are of interest. In our case, the assumption is the hopefulness level of the input sequence, namely $\sem{x,\exists O.\psi}$, and the guarantee is the satisfaction value of the specification in the generated computation, namely $\sem{x \otimes f(x),\psi}$. We assume that the desired relation between the assumption and the guarantee is given by a function $\comb:[0,1] \times [0,1] \to [0,1]$. For example, $\comb(A,G)=\max\{1-A,G\}$ captures  implication, and $\comb(A,G)=1-(A-G)$ measures the distance between the satisfaction value of the generated computation and the highest possible satisfaction value for the input sequence. When synthesizing, for example, a robot controller (e.g., vacuum cleaner) in a building, the doors to rooms are controlled by the environment, whereas the movement of the robot by the system. A measure of the performance of the robot has to take into an account both the number of ``hopeful rooms", namely these with an open  door, and the number of room cleaned. Note that  the satisfaction value of the assumption $\sem{x,\exists O.\psi}$ measures the performance of a good-enough {\em off-line\/} system. Thus, using a function $\comb(A,G)=\frac{G}{A}$, we can synthesize an {\em on-line\/} system with the best {\em competitive ratio\/} \cite{BE98} (see Example~\ref{xmp: elevator}). We show that the \ges-synthesis Safraless algorithm we suggested in the Boolean setting can be extended to handle \FLTL formulas in all the above approaches, thus the problem can be solved in doubly-exponential time.

	We continue with an analysis of the expected performance of the system. We do so by assuming a stochastic environment, with a known distribution on the input sequences. We introduce and study two measures for high-quality \ges-synthesis in a stochastic environment. In the first, termed \emph{expected \ges-synthesis}, all input sequences are sampled, yet the satisfaction value in each input sequence takes its hopefulness level into account, for example by a $\comb$ function as in the assume-guarantee setting. In the second, termed \emph{conditional expected \ges-synthesis}, only hopeful input sequences are sampled. For both approaches, our synthesis algorithm is based on the high-quality \FLTL synthesis algorithm of \cite{AK16}, which is based on an analysis of deterministic automata associated with the different satisfaction values of the \FLTL specification. Here too, the complexity stays doubly exponential. In addition, we extend the synthesized systems with guarantees for satisfaction and monitors indicating satisfaction in various satisfaction levels.

	\section{Preliminaries}	
	Consider two finite sets $I$ and $O$ of input and output signals, respectively. 
	For two words $x=i_0 \cdot i_1 \cdot i_2 \cdots \in (2^I)^\omega$ and $y = o_0 \cdot o_1 \cdot o_2 \cdots \in (2^I)^\omega$, we define $x \otimes y$ as the word in $(2^{I \cup O})^\omega$ obtained by merging $x$ and $y$. Thus, $x\otimes y = (i_0 \cup o_0)  \cdot (i_1 \cup o_1) \cdot (i_2 \cup o_2) \cdots$.
	The definition is similar for finite $x$ and $y$ of the same length. For a word $w \in (2^{I \cup O})^\omega$, we use $w_{|I}$ to denote the projection of $w$ on $I$. In particular, $(x\otimes y)_{|I}=x$.

	A {\em strategy\/} is a function $f:(2^I)^+ \rightarrow 2^O$. Intuitively, $f$ models the interaction of a system that generates in each moment in time a letter in $2^O$ with an environment that generates letters in $2^I$. For an input sequence $x=i_0 \cdot i_1 \cdot i_2 \cdots \in (2^I)^\omega$, we use $f(x)$ to denote the output sequence  $f(i_0) \cdot f(i_0 \cdot i_1) \cdot f(i_0 \cdot i_1 \cdot i_2) \cdots \in (2^O)^\omega$. Then, $x \otimes f(x) \in (2^{I \cup O})^\omega$ is the {\em computation\/} of $f$ on $x$. Note that the environment initiates the interaction, by inputting $i_0$. 
	Of special interest are {\em finite-state strategies}, induced by finite state transducers.  Formally, an {\em $I/O$-transducer} is $\T=\zug{I,O,S,s_0,M,\tau}$, where 
	$S$ is a finite set of states, $s_0 \in S$ is an initial state, $M: S \times 2^I \rightarrow S$ is a transition function, and $\tau:S \rightarrow 2^O$ is a labelling function. For $x=i_0 \cdot i_1 \cdot i_2 \cdots \in (2^I)^*$, let $M^*(x)$ be the state in $S$ that $\T$ reaches after reading $x$. Thus is, $M^*(\epsilon)=s_0$ and for every $j \geq 0$, we have that $M^*(i_0 \cdot i_1 \cdot i_2 \cdots i_j)=M(M^*(i_0 \cdot i_1 \cdot i_2 \cdots i_{j-1}),i_{j})$. Then, $\T$ induces the strategy $f_\T:(2^I)^+ \rightarrow 2^O$, where for every $x \in (2^I)^+$, we have that $f_\T(x)=\tau(M^*(x))$.
	We use $\T(x)$ and $x \otimes \T(x)$ to denote the output sequence and the computation of $\T$ on $x$, respectively, and talk about $\T$ realizing a specification, referring to the strategy $f_\T$.  
	
	We specify on-going behaviors of reactive systems using the {\em linear temporal logic\/} LTL \cite{Pnu81}. Formulas of LTL are
	constructed from a set $AP$ of atomic proposition using the usual
	Boolean operators and temporal operators like $\Alw$ (``always"), $\Ev$ (``eventually"), $\Next$ (``next time''), and
	$\Until$ (``until''). Each LTL formula $\psi$ defines a language $L(\psi)=\{ w : w \models \psi\} \subseteq (2^{AP})^\omega$. We also use {\em automata on infinite words\/} for specifying and reasoning about on-going behaviors. We use automata with different branching modes (nondeterministic, where some run has to be accepting; universal, where all runs have to be accepting; and deterministic, where there is a single run) and different acceptance conditions (B\"uchi, co-B\"uchi, and parity). We use the three letter acronyms NBW, UCW, DPW, and DFW, to refer to nondeterministic B\"uchi, universal co-B\"uchi, deterministic parity, and deterministic finite word automata, respectively. Given an LTL formula $\psi$ over $AP$, one can constructs an NBW $\A_{\psi}$ with at most ${2^{O(|\psi|)}}$ states such that $L(\A_{\psi})=L(\psi)$ {\rm \cite{VW94}}. Constructing an NBW for $\neg \psi$ and then dualizing it, results in a UCW for $L(\psi)$, also with at most ${2^{O(|\psi|)}}$ states. Determinization \cite{Saf88} then leads to a DPW for $L(\psi)$ with at at most ${2^{2^{O(|\psi|)}}}$ states and index ${2^{O(|\psi|)}}$. For full definitions of LTL, automata, and their relation, see \cite{Kup18}.
	
	Consider an LTL formula $\psi$ over $I \cup O$. 
	We say that $\psi$ is {\em realizable\/} if there is a finite-state strategy $f:(2^I)^+ \rightarrow 2^O$ such that for all $x \in (2^I)^\omega$, we have that $x \otimes f(x) \models \psi$. That is, the computation of $f$ on every input sequence satisfies $\psi$. 
	We say that a word  $x \in (2^I)^\omega$ is {\em hopeful\/} for $\psi$ if there is $y \in (2^O)^\omega$ such that $x \otimes y \models \psi$.
	Then, we say that $\psi$ is {\em good-enough realizable\/} (\ges-realizable, for short) if there is a finite-state strategy $f:(2^I)^+ \rightarrow 2^O$ such that for every $x \in (2^I)^\omega$ that is hopeful for $\psi$, we have that $x \otimes f(x) \models \psi$. That is, if there is some output sequence whose combination with $x$ satisfies $\psi$, then the computation of $f$ on $x$ satisfies $\psi$.  The LTL \ges-synthesis problem is then to decide whether a given LTL formula is \ges-realizable, and if so, to return a transducer that \ges-realizes it. 
	Clearly, every realizable specification is \ges-realizable -- by the same transducer. 
	We say that $\psi$ is {\em universally satisfiable\/} if all input sequences are hopeful for $\psi$. 
It is easy to see that for universally satisfiable specifications, realizability and \ges-realizability coincide. On the other hand, as demonstrated in Section~\ref{intro}, there are specifications that are not realizable and are \ges-realizable.
	
	\begin{example}
		\label{xmp: optimal synth vs lookahead}
		{\rm 
			%{\bf [Optimal Realizability vs Lookahead]}
			%Two common ``reasons'' for a specification not to be realizable are (1) the ability of the environment to violate the specification independently of the system's responses, and (2) a need for ``lookahead'' on the system's part (e.g., knowing what the next input would be).
			%At this point, the reader may suspect that optimal realizability is merely a workaround of (1), and that specifications that require lookahead are neither realizable nor optimally realizable.
			%As we now show, optimal realizability is a bit more subtle. 
			Let $I=\{p\}$ and $O=\{q\}$. Consider the specification $\psi= \Alw \Ev((\Next p) \wedge q) \wedge \Alw \Ev((\Next \neg p) \wedge  \neg q)$. Clearly, $\psi$ is not realizable, as an input sequence $x \in (2^I)^\omega$ is hopeful for $\psi$ iff $x \models \Alw \Ev p \wedge \Alw \Ev\neg p$. Since the system has to assign a value to $q$ before it knowns the value of $\Next p$, it seems that $\psi$ is also not \ges-realizable. As we show below, however, the specification $\psi$ is \ges-realizable. Intuitively, it follows from the fact that hopeful input sequences consists of alternating $p$-blocks and $(\neg p)$-blocks. Then, by outputting $\neg q$ in $p$-blocks and outputting $q$ in $(\neg p)$-blocks, the system guarantees that each last position in a $(\neg p)$-block satisfies  $q \wedge \Next p$ and each last position in a $p$-block satisfies $(\neg q) \wedge \Next p$.  Formally, $\psi$ is \ges-realized by the transducer $\T=\tup{\{p\},\{q\},\{s_0,s_1\},s_0,M,\tau}$, where $M(s_0,\emptyset)=M(s_1,\emptyset)=s_0$, $M(s_0,\{p\})=M(s_1,\{p\})=s_1$, $\tau(s_0)=\{q\}$,  and $\tau(s_1)=\emptyset$. \hfill \qed}	
		%	however, an optimally-realizing transducer can make use of the fact that $\psi$ has to be satisfied 
		%		
		%	However, this intuition is misleading, and $\psi$ is in fact optimally realizable. Indeed, consider the transducer $\T=\tup{I,O,\{s_0,s_1\},s_0,M,\tau}$ with the following transitions: $M(s_0,\emptyset)=M(s_1,\emptyset)=s_0$, and $M(s_0,\{p\})=M(s_1,\{p\})=s_1$. The labelling function is $\tau(s_0)=\{q\}$ and $\tau(s_1)=\emptyset$.
		%	
		%	Thus, $\T$ outputs $q$ until it sees $p$ in $s_1$, then proceeds to output $\neg q$ in $s_1$ until it sees $\neg p$, and goes back to $s_0$. If the environment outputs $p$ and $\neg p$ infinitely often, then the $\T$ will have outputted $q$ and $\neg q$, respectively, in the preceding steps, thus realizing the specification.	
	\end{example}

\stam{
\begin{remark}
\label{univsat}
An LTL formula $\psi$ over $I \cup O$ is {\em universally satisfiable\/} if all input sequences are hopeful for $\psi$. 
It is easy to see that for universally satisfiable specifications, realizability and \ges-realizability coincide.
%	\ges-synthesis may also remind the reader of synthesis with clairvoyance~\cite{KSS11}, where synthesis is relaxed by allowing the transducer some lookahead into the future. It is not hard to see that clairvoyance realizability and \ges-realizability are distinct notions, and they do not imply one another. 
\end{remark}
}	
	\section{LTL Good-Enough Synthesis}
	
	Recall that a strategy $f:(2^I)^+ \rightarrow 2^O$ \ges-realizes an LTL formula $\psi$ if its computations on all hopeful input sequences satisfy $\psi$. Thus, for every input sequence $x \in (2^I)^\omega$, either $x \otimes y \not \models \psi$ for all $y \in (2^O)^\omega$, or $x \otimes f(x) \models \psi$. The above suggests that algorithms for solving LTL \ges-synthesis involve existential and universal quantification over the behavior of output signals. 
	The logic EQLTL extends LTL by allowing existential quantification over atomic propositions \cite{SVW87}. We refer here to the case the atomic propositions are the signals in $I \cup O$, and the signals in $O$ are existentially quantified. Then, an EQLTL formula is of the form $\exists O.\psi$, and a computation $w \in (2^{I \cup O})^\omega$ satisfies $\exists O.\psi$ iff there is $y \in (2^O)^\omega$ such that $w_{|I} \otimes y \models \psi$. Dually, AQLTL extends LTL by allowing universal quantification over atomic propositions. We consider here formulas of the form $\forall O.\psi$, which are equivalent to $\neg \exists O.\neg \psi$. Indeed, a computation $w \in (2^{I \cup O})^\omega$ satisfies $\forall O.\psi$ iff for all $y \in (2^O)^\omega$, we have that $w_{|I} \otimes y \models \psi$. Note that in both the existential and universal cases, the $O$-component of $w$ is ignored. Accordingly, we sometimes interpret EQLTL and AQLTL formulas with respect to input sequences $x \in (2^I)^\omega$. Also note that both EQLTL and AQLTL increase the expressive power of LTL. For example, the EQLTL formula $\exists q. q \wedge \Next  \neg q \wedge \Alw (q \leftrightarrow \Next \Next q) \wedge \Alw(q \rightarrow p)$ states that $p$ holds in all even positions of the computation, which cannot be specified in LTL \cite{Wol81}. 
	
	\begin{theorem}
		\label{thm: ltl os 2exp}
		The LTL \ges-synthesis problem is 2EXPTIME-complete.
	\end{theorem}
	
	\begin{proof}
		We start with the upper bound. Given an LTL formula $\psi$ over $I \cup O$, we describe an algorithm that returns a transducer $\T$ that \ges-realizes $\psi$, or declares that no such transducer exists. 
		
		It is not hard to see that $\T$ \ges-realizes $\psi$ iff $\T$ realizes $\varphi=\psi \vee \forall O.\neg \psi$. Indeed, an input sequence $x \in   (2^I)^\omega$ is hopeful for $\psi$ iff $x \models \exists O. \psi$, and so the specification $\varphi$ requires all hopeful input sequences to satisfy $\psi$. 
		A naive construction of an NBW for $\varphi$ involves a universal projection of the signals in $O$ in an automaton for $\neg \psi$, and results in an NBW that is doubly exponential. In order to circumvent the extra exponent, we construct an NBW $\A_{\neg \varphi}$ for $\neg \varphi$, and then dualize it to get a UCW for $\varphi$, as follows. 
		
		Let $\A_{\neg \psi}$ be an NBW for $L(\neg \psi)$ and $\A_{\exists O. \psi}$ be an NBW for $L(\exists O. \psi)$. Thus, $\A_{\exists O. \psi}$ is obtained from an NBW $\A_\psi$ for $L(\psi)$ by existentially projecting its transitions on $2^I$. In more details, if $\A_\psi=\zug{2^{I \cup O}, Q,Q_0,\delta,\alpha}$, then 
		$\A_{\exists O. \psi}=\zug{2^{I \cup O}, Q,Q_0,\delta',\alpha}$, where for all $q \in Q$ and $i \cup o \in 2^{I \cup O}$, we have $\delta'(q,\sigma)=\bigcup_{o \in 2^O} \{\delta(q,(\sigma \cap I) \cup o)\}$. 
		
		Let $\A_{\neg \varphi}$ be an NBW for the intersection of $\A_{\neg \psi}$ and $\A_{\exists O. \psi}$. We can define $\A_{\neg \varphi}$ as the product of $\A_{\neg \psi}$ and $\A_{\exists O. \psi}$, possibly using the generalized B\"uchi acceptance condition (see Remark~\ref{efficient intersection}), thus its size is exponential in $\psi$. 
		The language of $\A_{\neg \varphi}$ is then $\{w \in (2^{I \cup O})^\omega: w \not \models \psi \mbox{ and } w \models \exists O.\psi\}$. We then solve usual synthesis for the complementing UCW. Its language is $\{w \in (2^{I \cup O})^\omega: w \models \psi \mbox{ or } w \models \forall O.\neg \psi\}$, as required. By \cite{KV05c}, the synthesis problem for UCW can be solved in EXPTIME, and we are done. 
		
		The lower bound follows from the 2EXPTIME-hardness of LTL realizability \cite{Ros92}. The hardness proof there constructs, given a 2EXPTIME Turing machine $M$, an LTL formula $\psi$ that is realizable iff $M$ accepts the empty tape. Since all input sequences are hopeful for $\psi$, realizability and \ges-realizability coincide, and we are done. 
		\qed
	\end{proof}
	
	Note that working with a UCW not only handles the universal quantification for free but also has the advantage of a Safraless synthesis algorithm -- no determinization and parity games are needed \cite{KV05c,KPV06}. Also note that the algorithm we suggest in the proof of Theorem~\ref{thm: ltl os 2exp} can be generalized to handle specifications that are arbitrary positive Boolean combinations of EQLTL formulas.

	\begin{remark}
		\label{efficient intersection}
		{\rm 
			{\bf [Products and Optimizations]}
			Throughout the paper, we construct products of automata whose state space is $2^{cl(\psi)}$, and states correspond to maximal consistent subsets of $cl(\psi)$, possibly in the scope of an existential quantifier of $O$. Accordingly, the product can be minimized to include only consistent pairs. 
			Also, since traditional-synthesis algorithms, in particular the Safraless algorithms we use, can handle automata with {\em generalized\/} B\"uchi and co-B\"uchi acceptance condition, we need only one copy of the product.  \hfill \qed}
	\end{remark}
	
	\begin{remark}
		\label{remark dual}
		{\rm 
			{\bf [Determinancy of the \ges-synthesis Game]}
			Determinancy of games implies that in traditional synthesis, a specification $\psi$ is not $I/O$-realizable iff $\neg \psi$ is $O/I$-realizable 
			This is useful, for example when we want to synthesize a transducer of a bounded size and proceed simultaneously, aiming to synthesize either a system transducer that realizes $\psi$ or an environment transducer that realizes $\neg \psi$ \cite{KV05c}. 
			For \ges-synthesis, simple dualization does not hold, but we do have determinancy in the sense that $(\exists O. \psi) \rightarrow \psi$ is not $I/O$-realizable iff $(\exists O. \psi) \wedge \neg \psi$ is $O/I$-realizable. Accordingly, $\psi$ is not \ges-realizable iff the environment has a strategy that generates, for each output sequence $y \in (2^O)^\omega$, a helpful input sequence $x \in (2^I)^\omega$ such that $x \otimes y \models \neg \psi$. In Appendix~\ref{app dual}, we formalize and study this duality further. \hfill \qed}
	\end{remark}

	\section{Guarantees in Good-Enough Synthesis}
	\label{subsec: guarantees in optimal}
	A drawback of \ges-synthesis is that we do not actually know whether the specification is satisfied. 
	In this section we describe two ways to address this drawback. 
	The first way goes beyond providing satisfaction information and enables the designer to partition the specification into to a {\em strong\/} component, which should be satisfied in all environments, and a {\em weak\/} component, which should be satisfied only in hopeful ones. The second way augments \ges-realizing transducers by flags, raised to indicate the status of the satisfaction.

	\subsection{\ges-Synthesis with a Guarantee}
	
	Recall that \ges-realizability is suitable especially in settings where we design a system that has to do its best in all environments. \ges-synthesis with a guarantee is suitable in settings where we want to make sure that some components of the specification are satisfied in all environment. Accordingly, a specification is an LTL formula $\psi = \psis \wedge \psiw$. When we {\em \ges-synthesize $\psiw$ with guarantee $\psis$}, we seek a transducer $\T$ that realizes $\psis$ and \ges-realizes $\psiw$. Thus, for all input sequences $x \in (2^I)^\omega$, we have that $x \otimes \T(x) \models \psis$, and if $x$ is hopeful for $\psiw$, then $x \otimes \T(x) \models \psis$.

	\begin{theorem}
		\label{thm: ltl oswg 2exp}
		The LTL \ges-synthesis with guarantee problem is 2EXPTIME-complete.
	\end{theorem}
	
	\begin{proof}
		Consider an LTL formula $\psi = \psis \wedge \psiw$ over $I \cup O$. It is not hard to see that a transducer $\T$ \ges-realizes $\psiw$ with guarantee $\psis$ iff $\T$ realizes $\varphi=\psis\wedge((\exists O. \psiw) \rightarrow \psiw)$. We can then construct a UCW $\A_\varphi$ for $L(\varphi)$ by dualizing an NBW for its negation $\neg \psis \vee ((\exists O.\psiw) \wedge \neg \psiw)$, which can be constructed using techniques similar to those in the proof of Theorem~\ref{thm: ltl os 2exp}. We then proceed with standard synthesis for $\A_\varphi$. Note that the approach is Safraless. Taking an empty  (that is, $\True$) guarantee, a lower bound follows from the 2EXPTIME-hardness of LTL \ges-synthesis.
		\qed
	\end{proof}
	
	\subsection{Flags by a \ges-Realizing Transducer}
	\label{subsec: flags}
	
	For a language $L \subseteq (2^{I \cup O})^\omega$ and a finite word $w \in (2^{I \cup O})^*$, let $L^w=\{w' \in (2^{I \cup O})^\omega : w \cdot w' \in L\}$. That is, $L^w$ is the language of suffixes of words in $L$ that have $w$ as a prefix. We say that a word $w \in (2^{I \cup O})^*$ is {\em green for\/} $L$ if $L^{w}$ is realizable. Then, a word $x \in (2^I)^*$ is {\em green for\/} $L$ if there is $y \in (2^O)^*$ such that $x \otimes y$ is green for $L$. When a system is lucky to interact with an environment that generates a green input sequence, we want the system to react in a way that generates a green prefix, and then realizes the specification. Formally, we say that a strategy $f:(2^I)^+ \rightarrow 2^O$ {\em green realizes\/} $L$ if for every $x \in (2^I)^+$, if $x$ is green for $L$, then $x \otimes f(x)$ is green for $L$.\footnote{Note that while the definition of green realization does not refer to $\epsilon$ directly, we have that $\epsilon$ is green iff $L$ is realizable, in which case all $x\in (2^I)^*$ are green.}
	\footnote{While synthesis corresponds to finding a winning strategy for the system, green synthesis can be viewed as a subgame-perfect best-response strategy, where the system does its best in every subgame, even if it loses the overall game.}
We say that a word $w \in (2^{I \cup O})^*$ is {\em light green for\/} $L$ if $L^{w}$ is universally satisfiable, thus 
	all input sequences are hopeful for $L^w$. 
	A word $x \in (2^I)^*$ is {\em light green for\/} $L$ if there is $y \in (2^O)^*$ such that $x \otimes y$ is light green for $L$. It is not hard to see that for \ges-realizable languages, green and light green coincide. Indeed, if $L$ is universally satisfiable and \ges-realizable, then $L$ is realizable.
	
	%Intuitively, even if a specification is not realizable, possibly also not realizable, if the system is lucky to interact with an environment that generate a green input sequence, then the system may respond in a way that would guarantee realizability. Formally, we say that a language $L \subseteq (2^{I \cup O})^\omega$ is {\em green realizable} if there is a strategy $f:(2^I)^* \rightarrow 2^O$ such that for every $x \in (2^I)^*$, if $x$ is green for $L$, then $x \otimes f(x)$ is green for $L$.
	% As we argue below, optimal realizability is strictly stronger than green realizability.
	
	\begin{theorem}
		\label{thm: optimal real implies flags}
		\ges-realizability is strictly stronger than green realizability.  
	\end{theorem}
	
	\begin{proof}
		We first prove that every strategy $f:(2^I)^+ \rightarrow 2^O$ that \ges-realizes a specification $\psi$ also green realizes $\psi$. Consider $x \in (2^I)^+$ that is green for $\psi$. By definition, there is $y \in (2^O)^+$ such that $L^{x \otimes y}$ is realizable. Then, for every $x' \in (2^I)^\omega$, there is $y' \in (2^O)^\omega$ such that $x' \otimes y'$ in $L^{x \otimes y}$. Hence, for every $x' \in (2^I)^\omega$, we have that $x \cdot x'$ is hopeful. Therefore, as $f$ \ges-realizes $\psi$, we have that $(x \cdot x') \otimes f(x \cdot x') \models \psi$. Thus, $x \otimes f(x)$ is green, and so $f$ green realizes $\psi$. 
		
		We continue and describe a specification that is green realizable and not \ges-realizable. Let $I=\{p\}$ and $O=\{q\}$. 
		Consider the specification $\psi= \Alw ((\Next p) \leftrightarrow q)$. Clearly, $\psi$ is not realizable, as the system has to commit a value for $q$ before a value for $Xp$ is known. Likewise, no word $w\in (2^{I\cup O})^*$ is green for $\psi$, and so no finite input sequence $x \in  (2^I)^*$ is green for $\psi$. Hence, 
		every strategy (vacuously) green realizes $\psi$. On the other hand, for every input sequences $x \in (2^I)^\omega$ there is an output sequence $y \in (2^O)^\omega$ such that $x \otimes y \models \psi$. Thus, all input sequences are hopeful for $\psi$. Thus, synthesis and \ges-synthesis coincide 
		for $\psi$, which is not \ges-realizable. 
		\hfill \qed
	\end{proof}

	Theorem~\ref{thm: optimal real implies flags} brings with it two good news. 
	The first is that a \ges-realizing transducer has the desired property of being also green realizing.
	The second has to do with our goal of providing the user with information about the satisfaction status, in particular raising a green flag whenever a green prefix is detected. By Theorem~\ref{thm: optimal real implies flags}, such a flag indicates that the computation generated by our \ges-realizing transducer satisfies the specification. 
	A naive way to detect green prefixes for a specification $\psi$ is to solve the synthesis problem for $\psi$ by solving a game on top of a DPW $\D_\psi$ for $\psi$. The winning positions in the game are states in $\D_\psi$. By defining them as accepting states, we can obtain from $\D_\psi$ a DFW for green prefixes. 
	Then, we run this DFW in parallel with the \ges-realizing transducer, and raise the green flag whenever a green prefix is detected. 
	This, however, requires a generation of $\D_\psi$ and a solution of parity games. Below we describe a much simpler way, which makes use of the fact that our transducer \ges-realizes the specification. 

	Recall that if $L$ is universally satisfiable and \ges-realizable, then $L$ is realizable.
	Accordingly, given a transducer $\T$ that \ges-realizes $\psi$, we can augment it with green flags by running in parallel a DFW that detects light-green prefixes. As we argue below, constructing such a DFW only requires an application of the subset construction on top of an NBW for the existential projection of $\psi$ on $2^I$. 
	
	\begin{lemma}
		\label{lem: light green DFAs}
		Given an LTL formula $\psi$ over $I \cup O$, we can construct a DFA $\S$ of size $2^{2^{O(|\psi|)}}$ such that $L(\S)=\{x\in (2^{I})^*: x \mbox{ is light green for $L(\psi)$}\}$. 
	\end{lemma}
	\begin{proof}
		Let $\A_\psi = \zug{2^{I \cup O},Q,\delta,Q_0,\alpha}$ be an NBW for $L(\psi)$, and let $\B_\psi = \langle 2^{I},Q$, $\delta'$, $Q_0, \alpha \rangle$ be its existential projection on $2^I$. Thus, for every $q \in Q$ and $i \in 2^I$, we have $\delta'(q,i)=\bigcup_{o \in 2^O}\delta(q,i \cup o)$. 
		We define the DFW $\S = \zug{2^I,2^Q,M,\{Q_0\},F}$, where $M$ follows the subset construction of $\B_\psi$: for every $S \in 2^Q$ and $i \in 2^I$, we have $M(S,i)= \bigcup_{s\in S} \delta'(s,i)$. Then, $F=\{S \in 2^Q: L(\B_\psi^S) = (2^I)^\omega\}$. Observe that $\S$ rejects $x \in (2^I)^*$ iff there is $x' \in (2^I)^\omega$ such that for all $y \in (2^O)^*$ and $y' \in (2^O)^\omega$, no state in $\delta(Q_0,x \otimes y)$ accepts $x' \otimes y'$. Thus, $\S$ rejects $x$ iff $x$ is not light green, and accepts it otherwise. Note that the definition of $F$ involves universality checking, possibly via complementation, yet no determinization is required, and the size of $\S$ is $2^{2^{O(|\psi|)}}$.		\hfill \qed
	\end{proof}
	
	Note that once we reach an accepting state in $\S$, we can make it an accepting loop. Indeed, once a green prefix is detected, then all prefixes that extend it are green. Accordingly, once the green flag is raised, it stays up. 
	Also note that if an input sequence is not hopeful for $\psi$, then none of its prefixes is light green for $\psi$. The converse, however, is not true: an input sequence may be hopeful and still have no light green prefixes. For example, taking $I=\{p\}$, the input sequence $\{p\}^\omega$ is hopeful for $\Alw p$, yet none of its prefixes is green light, as it can be extended to an input sequence with $\neg p$. 
	
	Green flags provide information about satisfaction. Two additional flags of interest are related to safety and co-safety properties:
	\begin{itemize}
		\item
		A word $w \in (2^{I \cup O})^*$ is {\em red for\/} $L$ if $L^{w}=\emptyset$.
		A word $x \in (2^I)^*$ is {\em red for\/} $L$ if for all $y \in (2^O)^*$, we have that $x \otimes y$ is red for $L$.
		Thus, when the environment generates $x$, then no matter how the system responds, $L$ is not satisfied. 
		\item
		a word $w \in (2^{I \cup O})^*$ is {\em blue for\/} $L$ when $L^{w}=(2^{I \cup O})^\omega$, and then define  
		a word $x \in (2^I)^*$ as {\em blue for\/} $L$ if there is $y \in (2^O)^*$ such that $x \otimes y$ is blue for $L$. 
		Thus, when the environment generates $x$, the system can respond in a way that guarantees satisfaction no matter how the interaction continues. 
	\end{itemize}
	
	A monitor that detects red and blue prefixes for $L$ can be added to a transducer that \ges-realizes $L$. As has been the case with the monitor for green prefixes, its construction is based on applying the subset construction on an NBW for $L$ \cite{KV01d}. Also, as in the green case, once a red or blue flag is raised, it stays up. In a way analogous to green realizability, we seek a transducer that \ges-realizes the specification and generates a red prefix only if all interactions generate a red prefix, and generates a blue prefix whenever this is possible. In Appendix~\ref{redblue}, we show that while \ges-realization implies {\em red realization}, it may conflict with {\em blue realization}.

	\section{High-Quality Good-Enough Synthesis}
	
	\ges-synthesis is of special interest when the satisfaction value of the specification is multi-valued, and we want 
	%short
	to synthesize high-quality systems.
	%the synthesized system to achieve the highest possible satisfaction value for each input sequence. 
	We start by defining the multi-valued logic \FLTL, which is our multi-valued specification formalism. We then study \FLTL \ges-synthesis, first in a worst-case approach, where the satisfaction value of a transducer is the satisfaction value of its computation with the lowest satisfaction value, and then in a stochastic approach, where it is the expected satisfaction value, given a distribution of the inputs.

	\subsection{The logic \FLTL}
	
	Let $AP$ be a set of Boolean atomic propositions and let $\F\subseteq\{f:[0,1]^k\to [0,1]:k\in \N\}$ be a set of {\em quality operators}. An \FLTL formula is one of the following:
	\begin{compactitem}
		\item $\True$, $\False$, or $p$, for $p\in AP$.
		\item $f(\psi_1,...,\psi_k)$, $\Next\psi_1$, or $\psi_1\Until \psi_2$, for \FLTL formulas $\psi_1,\ldots,\psi_k$ and a function $f\in \F$.
	\end{compactitem}
	The semantics of \FLTL formulas is defined with respect to infinite computations over $AP$. For a computation $w=w_0,w_1,\ldots \in (2^{AP})^\omega$ and position $j \geq 0$, we use $w^j$ to denote the suffix $w_j,w_{j+1},\ldots$. The semantics maps a computation $w$ and an \FLTL formula $\psi$ to the {\em satisfaction value\/} of $\psi$ in $w$, denoted $\sem{w,\psi}$. The satisfaction value is in $[0,1]$ and is defined inductively as follows. \footnote{The observant reader may be concerned by our use of $\max$ and $\min$ where $\sup$ and $\inf$ are in order. In Theorem~\ref{from abk} we state that there are only finitely many satisfaction values for a formula, thus the semantics is well defined.}
	\begin{itemize}
		\item 
		$\sem{w,\True}=1$ and  $\sem{w,\False}=0$.
		\item
		For $p \in AP$, we have that $\sem{w, p}=1$ if $p\in w_0$, and $\sem{w, p}=0$ if $p\not \in w_0$. 
		\item
		$\sem{w,f(\psi_1,...,\psi_k)}=f(\sem{w,\psi_1},...,\sem{w,\psi_k})$.
		\item
		$\sem{w, \Next \psi_1}=\sem{w^1, \psi_1}$.
		\item
		$\sem{w, \psi_1 \Until \psi_2}=\max\limits_{i\geq 0} \{ \min \{\sem{w^i,\psi_2}, \min\limits_{0\leq j < i}\sem{w^j,\psi_1} \} \}$. 
	\end{itemize}
	
	The logic \LTL can be viewed as \FLTL for $\F$ that models the usual Boolean operators. In particular, the only possible satisfaction values are $0$ and $1$. For simplicity, we abbreviate common functions as described below. In addition, we introduce notations for two useful quality operators, namely factoring and weighted average. Let $x,y,\lambda \in[0,1]$. Then, 
	\begin{center}
		\begin{tabular}{lll}
			%space
			$\bullet ~~ \neg x=1-x$ \,  & $\bullet ~~ x\vee y=\max\{x,y\}$ \,  & $\bullet ~~ x\wedge y=\min\{x,y\}$ \\
			$\bullet ~~ x\rightarrow y=\max\{1-x,y\}$  \, & $\bullet ~~ \factorU_\lambda x=\lambda\cdot x$  \, & $\bullet ~~ x\avg{\lambda}y=\lambda \cdot x+(1-\lambda)\cdot y$
		\end{tabular}
	\end{center}
	
	\begin{example}
		\label{xpl: fltl formula}
		{\rm Consider a scheduler that receives requests and generates grants, and consider the \FLTL formula
			$\psi=\psi_1\avg{\frac34} \psi_2$, with $\psi_1=\Alw({\it req} \rightarrow ({\it grant}\vee \factorU_{\frac{2}{3}}\Next{\it grant}))$ and $\psi_2=\Ev {\it req}$. The satisfaction value of the formula $\psi_1$ is $1$ if every request is granted immediately. If the grant is given with a delay, the satisfaction value reduces to $\frac{2}{3}$. In addition, the weighted average with $\psi_2$ implies that $\psi_1$ contributes up to $\frac34$ of the satisfaction value of $\psi$: if there are no requests, and thus $\psi_2$ is violated, then the satisfaction of $\psi_1$ is vacuous, and the satisfaction value of $\psi$ is penalized by $\frac14$. The example demonstrates how \FLTL can conveniently prioritize different scenarios, as well as embody vacuity considerations in the formula. 
			\hfill \qed
		}
	\end{example}
	
	The realizability problem for \FLTL is an optimization problem: For an \FLTL specification $\psi$ and a transducer $\T$, we define the satisfaction value of $\psi$ in $\T$, denoted $\sem{\T,\psi}$, by $\min \{\sem{x\otimes \T(x),\psi}: x \in (2^I)^\omega\}$, namely the satisfaction value of $\psi$ in the worst-case. Then, the synthesis problem is to find, given $\psi$, a transducer that maximizes its satisfaction value. Moving to a decision problem, given $\psi$ and a threshold value $v\in [0,1]$, we say that $\psi$ is \emph{$v$-realizable} if there exists a transducer $\T$ such that $\sem{\T,\psi} \geq v$, and the synthesis problem is to find, given $\psi$ and $v$, a transducer $\T$ that $v$-realizes $\psi$. 
	%We lift realizability to the quantitative setting using thresholds: for an \FLTL specification $\psi$ over $I\cup O$ and a threshold $v\in [0,1]$, we say that $\psi$ is \emph{$v$-realizable} if there exists a strategy $f:(2^I)^+\to 2^O$ such that for every $x\in \Io$? we have that $\sem{x\otimes f(x),\psi}\ge v$. 

	For an \FLTL formula $\psi$, let $V(\psi)$ be the set of possible satisfaction values of $\psi$ in arbitrary computations. Thus, $V(\psi)=\{\sem{w,\psi} \ : \  w\in (2^{AP})^{\omega}\}$. 
	\begin{theorem}
		\label{from abk}{\rm \cite{ABK16}}
		Consider an \FLTL formula $\psi$.
		\begin{itemize}
			\item
			$|V(\psi)|\le 2^{|\psi|}$.
			\item
			For every predicate $P \subseteq [0,1]$, there exists an NBW $\A^P_{\psi}$ such that $L(\A^P_{\psi})=\{ w :\sem{w,\psi} \in P\}$. Furthermore, $\A^P_{\psi}$ has at most ${2^{O(|\psi|^2)}}$ states {\rm \cite{ABK16}}.
		\end{itemize}
	\end{theorem}

	As with LTL, we define the existential and universal extensions \EQFLTL and \AQFLTL of \FLTL. Here too, we consider the case $AP=I \cup O$, with the signals in $O$ being quantified. Then, $\sem{w,\exists O.\psi}=\max_{y \in (2^O)^\omega} \{ \sem{w_{|I} \otimes y,\psi}\}$ and $\sem{w,\forall O.\psi}=\min_{y \in (2^O)^\omega} \{ \sem{w_{|I} \otimes y,\psi}\}$. 
	
	\begin{remark}
		\label{rmk: sem of exists forall}{\rm 
			{\bf [On the Semantics of \EQFLTL]}
			It is tempting to interpret an expression like $\sem{w,\exists O.\psi} \leq v$ as ``there exists an output sequence $y$ such that $\sem{w_I\otimes y,\psi} \leq v$''. By the semantics of $\exists O.\psi$, however, $\sem{w,\exists O.\psi} \leq v$ actually means that $\max_{y\in (2^O)^\omega}\sem{w_I\otimes y,\psi} \leq v$. Thus, the correct interpretation is  ``for all output sequences $y$, we have that $\sem{w_I\otimes y,\psi} \leq v$".
			\hfill \qed
		}
	\end{remark}

	\subsection{\FLTL \ges-Synthesis}
	\label{sec: FLTL optimal synt}
	\label{fltl os}
	For a value $v \in [0,1]$, we say that $x$ is {\em $v$-hopeful for $\psi$\/} if there is $y \in (2^O)^\omega$ such that $\sem{x \otimes y,\psi} \geq v$.
	We study two variants of \FLTL \ges-synthesis:
	\begin{itemize}
		\item
		In {\em \FLTL \ges-synthesis with a threshold}, the input is an \FLTL formula $\psi$ and a value $v \in [0,1]$, and the goal is to generate a transducer whose computation on every input sequence that is $v$-hopeful has satisfaction value at least $v$. Formally, a function $f:(2^I)^+ \rightarrow 2^O$ \ges-realizes $\psi$ with threshold $v$ if for every $x \in (2^I)^\omega$, if $x$ is $v$-hopeful, then $\sem{x \otimes f(x), \psi} \geq v$. 
		\item
		In {\em \FLTL \ges-synthesis}, the input is an \FLTL formula $\psi$, and the goal is to generate a transducer whose computation on every input sequence has the highest possible satisfaction value for this input sequence.  
		Formally, a function $f:(2^I)^+ \rightarrow 2^O$ \ges-realizes $\psi$ if for every $x \in (2^I)^\omega$ and value $v \in [0,1]$, if $x$ is $v$-hopeful, then $\sem{x \otimes f(x), \psi} \geq v$. 
	\end{itemize}
	
	In the Boolean case, the two variants coincide, taking $v=1$. Indeed, then, for every $x \in (2^I)^\omega$, if $x$ is hopeful, then $x \otimes f(x)$ has to satisfy $\psi$. We note that \ges-realization with a threshold is not monotone, in the sense that decreasing the threshold need not lead to \ges-realization. Indeed, the lower is the threshold $v$, the more input sequences are $v$-helpful (see Example~\ref{v-helpful example}). Accordingly, we do not search for a maximal threshold,  and rather may ask about a desired threshold or about \ges-synthesis without a threshold.

	Solving the \ges-synthesis problem, a naive combination of the automata construction of Theorem~\ref{from abk} with the projection technique of Theorem~\ref{thm: ltl os 2exp}, corresponds to an erroneous semantics of \EQFLTL, as noted in Remark~\ref{rmk: sem of exists forall}. Before describing our construction, it is helpful to state the correct (perhaps less intuitive) interpretation of existential and universal quantification in the quantitative setting:
	
	\begin{lemma}
		\label{existsforall}
		For every \FLTL formula $\psi$ and an input sequence $x \in (2^I)^\omega$, we have that $\sem{x,\exists O.\psi}=1-\sem{x,\forall O.\neg \psi}$. Accordingly, for every value $v \in [0,1]$, we have that $\sem{x,\exists O.\psi} < v$ iff $\sem{x,\forall O.\neg \psi} > 1-v$. 
	\end{lemma}
	
	\begin{proof}
		By definition, 
		$\sem{x,\exists O.\psi}=\max_{y \in (2^O)^\omega} \sem{x \otimes y,\psi}=1-\min_{y \in (2^O)^\omega} 1- \sem{x \otimes y,\psi}=1-\min_{y \in (2^O)^\omega} \sem{x \otimes y,\neg \psi}=1-\sem{x,\forall O.\neg \psi}$. 
		Then, 
		$\sem{x,\exists O.\psi} < v$ iff $1-\sem{x,\exists O.\psi} > 1-v$ iff $\sem{x,\forall O.\neg \psi}) > 1-v$.
		\qed
	\end{proof}	
	
	Consider an \FLTL formula $\psi$, a value $v \in [0,1]$, and an input sequence $x \in (2^I)^\omega$. Recall that $x$ is $v$-hopeful for $\psi$ if there is $y \in (2^O)^\omega$ such that $\sem{x \otimes y,\psi} \geq v$. Equivalently, $\sem{x,\exists O.\psi} \geq v$. 
	Indeed, $\sem{x,\exists O.\psi} = \max_{y \in (2^O)^\omega} \sem{x \otimes y,\psi}$, which is greater or equal to $v$ iff there is $y \in (2^O)^\omega$ such that $\sem{x \otimes y,\psi} \geq v$. 
	Hence, $x$ is not $v$-hopeful for $\psi$ if $\sem{x,\exists O.\psi} < v$. Equivalently, by Lemma~\ref{existsforall}, $\sem{x,\forall O.\neg \psi} > 1-v$. 
	Accordingly, for a strategy $f:(2^I)^+ \rightarrow 2^O$, an input sequence $x \in (2^I)^\omega$, and a value $v \in [0,1]$, we say that $f$ is {\em $v$-good for $x$ with respect to $\psi$}, if $\sem{x \otimes f(x),\psi} \geq v$ or $\sem{x,\forall O.\neg \psi} > 1-v$.

	\begin{example}
		\label{v-helpful example}
		Let $I=\{p\}$ and $O=\{q\}$. Consider the \FLTL formula $\psi=(\factorU_{\frac14} p \vee \factorU_{\frac12}q)$. Checking for which values $v$ a strategy $f$ is $v$-good for $x$ with respect to $\psi$, we examine whether $\sem{x \otimes f(x), \factorU_{\frac14} p \vee \factorU_{\frac12}q} \geq v$ or $\sem{x, \forall q.\neg( \factorU_{\frac14} p \vee \factorU_{\frac12}q)} >1-v$.
		Since $\psi$ refers only to the first position in the computation, it is enough to examine $x_0$ and $f(x_0)$. 
		For example, if $x_0=\emptyset$ and $f(x_0)=\emptyset$, then
		$\sem{x \otimes f(x), \factorU_{\frac14} p \vee \factorU_{\frac12}q} = 0$, $\sem{x,\exists q.\factorU_{\frac14} p \vee \factorU_{\frac12}q}=\max\{0,\frac12\}=\frac12$, and 
		$\sem{x, \forall q. \neg(\factorU_{\frac14} p \vee \factorU_{\frac12}q)}=\min\{1,1-\frac12\}=\frac12$.
		Hence, $f$ is $v$-good for $x$ with respect to $\psi$ if $v = 0$ or $v > \frac12$, thus $v \in \{0\} \cup (\frac12,1]$. Similarly, we have the following (see Appendix~\ref{apx:example helpful} for a detailed analysis).
		\begin{itemize}
			\item If $x_0=\emptyset$ and $f(x_0)=\{q\}$ then $f$ is $v$-good for $x$ when $v\in[0,1]$.
			\item If $x_0=\{p\}$ and $f(x_0)=\emptyset$ then $f$ is $v$-good for $x$ when $v\in [0,\frac14]\cup (\frac12,1]$.
			\item If $x_0=\{p\}$ and$f(x_0)=\{q\}$ then $f$ is $v$-good for $x$ when $v\in [0,1]$.
		\end{itemize}
	\end{example}

	\begin{theorem}
		\label{fltl with v}
		\label{thm: FLTL os with threshold 2exp}
		The \FLTL \ges-synthesis with threshold problem is 2EXPTIME-complete.
	\end{theorem}
	
	\begin{proof}
		We start with the upper bound, and show we can adjust the upper bound described in the proof of Theorem~\ref{thm: ltl os 2exp} to the multi-valued setting. Given an \FLTL formula $\psi$ over $I \cup O$ and a threshold $v \in [0,1]$, we describe an algorithm that returns a transducer $\T$ that \ges-realizes $\psi$ with threshold $v$, or declares that no such transducer exists. 
		
		By definition, we have that $\T$ \ges-realizes $\psi$ with threshold $v$ if for every input sequence $x$, we have that $f_\T$ is $v$-good for $x$ with respect to $\psi$. Thus, $\sem{x \otimes f_\T(x),\psi} \geq v$ or $\sem{x,\forall O.\neg \psi} > 1-v$. 
		We construct a UCW whose language is $\{w \in (2^{I \cup O})^\omega: \sem{w, \psi} \geq v \mbox{ or } \sem{w,\forall O.\neg \psi} > 1-v\}$.
		
		Let $\A^{<v}_{\psi}$ be an NBW for $\{w: \sem{w, \psi} < v\}$ and $\A^{\ge v}_{\exists O. \psi}$ be an NBW for $\{w: \sem{w, \exists O. \psi} \geq v\}$. Thus, $\A^{\ge v}_{\exists O. \psi}$ is obtained from an NBW $\A^{\ge v}_\psi$ for $\{w: \sem{w, \psi} \geq v\}$ by existentially projecting its transitions on $2^I$. By Theorem~\ref{from abk}, both $\A^{< v}_{\psi}$ and $\A^{\ge v}_{\exists O. \psi}$ are of size exponential in $\psi$. 
		
		Let $\B^v_\psi$ be an NBW for the intersection of $\A^{<v}_{\psi}$ and $\A^{\ge v}_{\exists O. \psi}$. The language of $B^v_\psi$ is then $\{w \in (2^{I \cup O})^\omega: \sem{w, \psi} < v \mbox{ and }  \sem{w, \exists O. \psi} \geq v\}$. We then solve usual synthesis for the complementing UCW, whose language is $\{w \in (2^{I \cup O})^\omega: \sem{w, \psi} \geq v \mbox{ or } \sem{w,\forall O.\neg \psi} > 1-v\}$, as required. By \cite{KV05c}, the synthesis problem for UCW can be solved in EXPTIME.
		
		The lower bound follows from the 2EXPTIME-hardness of LTL \ges-realizability.  
		\qed
	\end{proof}
	
	\begin{theorem}
		\label{fltl without v}
		The \FLTL \ges-synthesis problem is 2EXPTIME-complete.
	\end{theorem}

	\begin{proof}
		We start with the upper bound. Given an \FLTL specification $\psi$ over $I \cup O$, we describe an algorithm that returns a transducer $\T$ that \ges-realizes $\psi$ or declares that no such transducer exists. 
		
		As discussed above, a transducer $\T$ \ges-realizes $\psi$ iff for every input sequence $x \in (2^I)^\omega$ and value $v \in [0,1]$, we have that $f_\T$ is $v$-good for $x$ with respect to $\psi$.  Accordingly, we construct a UCW whose language is $\bigcap_{v \in V(\psi)} \{w \in (2^{I \cup O})^\omega: \sem{w, \psi}\geq v \mbox{ or } \sem{w,\forall O.\neg \psi} > 1-v\}$.
		
		For $v \in V(\psi)$, let $\B^v_{\psi}$ be an NBW for $\{w: \sem{w, \neg \psi} \geq v \mbox{ and } \sem{w, \exists O. \psi} \geq v\}$, as constructed in the proof of Theorem~\ref{fltl with v}, and let $\B$ be the union of $\B^v_{\psi}$ for all $v \in V(\psi)$. By Theorem~\ref{from abk}, the size of $V(\psi)$ is exponential in $\psi$, and thus so is the size of $\B$. We then solve usual synthesis for the complementing UCW, whose language is as required. By \cite{KV05c}, the synthesis problem for UCW can be solved in EXPTIME.
		The lower bound follows from the 2EXPTIME-hardness of LTL \ges-realizability.  
		\qed
	\end{proof}
	
	\begin{remark}\label{factor hope}{\bf [Tuning Hope Down]}
		The quantitative setting allows the designer to tune down ``satisfaction by hoplessness": rather than synthesizing  $\psi \vee \forall O. \neg \psi$, we can have a factor $\lambda$ and synthesize $\psi \vee \factorU_\lambda \forall O. \neg \psi$. In Section~\ref{fltl ag os} below we study additional ways to refer to hopefulness levels. 
	\end{remark}
	
	\subsection{\FLTL Assume-Guarantee  \ges-Synthesis}
	\label{subsec: assume guarantee}
	\label{fltl ag os}
	
	In Section~\ref{fltl os}, we seek a transducer $\T$ such that for a given or for all values $v \in [0,1]$ and input sequences $x \in (2^I)^\omega$, if $\sem{x,\exists O.\psi} \geq v$ then $\sem{x \otimes \T(x), \psi} \geq v$. In this section we measure the quality of a transducer $\T$ by analyzing richer relations between $\sem{x,\exists O.\psi}$ and $\sem{x \otimes \T(x), \psi}$. % for the different input sequences $x \in (2^I)^\omega$. 
	The setting has the flavor of quantitative assume-guarantee synthesis \cite{AKRV17}. There, the specification consists of a multi-valued assumption $A$, which in our case is $\exists O.\psi$, and a multi-valued guarantee $G$, which is our case is $\psi$.  
	
	There are different ways to analyze the relation between $\sem{x,\exists O.\psi}$ and $\sem{x \otimes \T(x), \psi}$.  To this end, we assume that we are given a function $\comb:[0,1] \times [0,1] \to [0,1]$ that given the satisfaction values of $\exists O. \psi$ and of $\psi$, outputs a combined satisfaction value. We assume that $\comb$ is decreasing in the first component and increasing in the second component. 
	This corresponds to the intuition that a lower satisfaction value of $\exists O. \psi$ and a higher satisfaction value of $\psi$ both yield a higher overall score.
	Also, since $\sem{x,\exists O.\psi} \geq \sem{x \otimes \T(x), \psi}$ for all $x \in (2^I)^\omega$, we assume that the first component is greater than or equal to the second. 
	Finally, we require $\comb$ to be efficiently computed. 
	Some natural $\comb$ functions include:
	\begin{itemize}
		\item The quantitative implication function: $\comb(A,G)=\max\{1-A,G\}$. This captures the quantitative notion of the implication $(\exists O.\psi) \rightarrow \psi$.
		\item The (negated) difference function: $\comb(A,G)=1-(A-G)$. This captures how far the satisfaction value for the given computation is from the 
		%optimal 
		best 
		satisfaction value. Since $A\ge G$, the range of the function is indeed $[0,1]$.
		\item The ratio function, given by some normalization to $[0,1]$ of the function $\comb(A,G)=\frac{G}{A}$, which captures the ``relative success'' with respect to the best possible satisfaction value. 
	\end{itemize} 
	
	The choice of an appropriate $\comb$ function depends on the setting. 
		Implication is in order when harsh environments may outweigh the actual performance of the system. For example, if our specification measures the uptime of a server in a cluster, then environments that cause very frequent power failures render the server unusable, as the overhead of reconnecting it outweighs its usefulness. In such a case, being shut down is better than continuously trying to reconnect, and so we give a higher satisfaction value for the server being down, which depends only on the environment.
	Then, as demonstrated with the cleaning robot in Section~\ref{intro}, the difference and ratio functions are fairly natural when measuring ``realization of potential''. We now describe a more detailed example when these measures are in order. 
	\begin{example}
		\label{xmp: elevator}
		{\rm 
			Consider a controller for an elevator in an $n$-floor building. In each moment in time, the environment sends to the controller requests, by means of a truth assignment to $I=\{1,\ldots,n\}$, indicating the subset of floors in which the elevator is requested. Then, the controller assigns values to $O=\{\textit{up}, \textit {down}\}$, directing the elevator to go up, go down, or stay. The satisfaction value of the specification $\psi$ reflects the waiting time of the request with the slowest response: it is $0$ when this time is more than $2n$, and is $1$ when the slowest request is granted immediately. 
			%In the initial configuration, the elevator is in floor $1$ and there are no pending requests. 
			Sure enough, there is no controller that attains satisfaction value $1$ on all input sequences, and so $\psi$ is not realizable with satisfaction value $1$. Also, adding assumptions about the behavior of the environment is not of much interest. Using AG \ges-realizability, we can synthesize a controller that behaves 
			%in an optimal way. 
			as well as possible.
			For example, using the difference function, we measure the performance of the controller on an input sequence $x \in (2^I)^\omega$ with respect to the best possible performance on $x$. Note that such a best performance needs a look-ahead on requests yet to come, which is indeed the satisfaction value of $\exists O.\psi$ in $x$. Thus, the assumption $\sem{x,\exists O.\psi}$ actually gives us the performance of a good-enough {\em off-line\/} controller. Accordingly, using the ratio  function, we can synthesize a system with the best {\em competitive ratio} for an on-line interaction \cite{BE98}. 
			\hfill \qed}
	\end{example}	

%of stam
	
	Given an \FLTL formula $\psi$ and a function $\comb$, we define the \emph{\ges-AG-realization value\/} of $\psi$ in a transducer $\T$ by $\min\{\comb(\sem{x,\exists O.\psi}, \sem{x\otimes\T(x),\psi}) : x\in (2^I)^\omega\}$.
	Then, our goal in \emph{AG \ges-realizability} is to find, given an \FLTL formula $\psi$ and a function $\comb$, the maximal value $v \in [0,1]$ such that there exists a transducer $\T$ whose AG \ges-realization value of $\psi$ is $v$. The {\em AG \ges-synthesis\/} problem is then to find such a transducer.
	
	We start by solving the decision version of AG \ges-realizability.
	
	\begin{theorem}
		\label{thm: quant opt synth decision}
		The problem of deciding, given an \FLTL formula $\psi$, a function $\comb$, and a threshold $v\in [0,1]$, whether there exists a transducer $\T$ whose AG \ges-realization value of $\psi$ is $v$, is 2EXPTIME-complete.
	\end{theorem}
	\begin{proof}
		Recall that $V(\psi)$ is the set of possible satisfaction values of $\psi$ (and hence of $\exists O. \psi$), and that by Theorem~\ref{from abk}, we have that $|V(\psi)|\le 2^{|\psi|}$. Let $G_v=\{\zug{v_1,v_2}\in V(\psi)\times V(\psi): \comb(v_1,v_2)\ge v\}$. Intuitively, $G$ is the set of satisfaction-value pairs $\zug{\sem{w,\exists O.\psi},\sem{w,\psi}}$ that are allowed to be generated by a transducer whose AG \ges-realization value of $\psi$ is at least $v$.  By definition,  AG \ges-realization of $\psi$ with value $v$ coincides with realization of the language $L_v=\{w\in \IOo: \comb(\sem{w,\exists O. \psi},\sem{w,\psi})\ge v \}$. By the monotonicity assumption on $\comb$, for every $\zug{v_1,v_2}\in G_v$, we have that $\zug{v'_1,v'_2}\in G$ for every $v'_1\le v_1$ and $v'_2\ge v_2$. Hence, we can write 
		\[L_v=\bigcup_{\zug{v_1,v_2}\in G_v}\{w\in \IOo: \sem{w,\exists O.\psi}\le v_1 \mbox{ and } \sem{w,\psi}\ge v_2 \},\] and proceed to construct an NBW for $L_v$ by taking the union of NBWs $\A_{v_1,v_2}$ for all $\zug{v_1,v_2}\in G_v$, each of which is the product of NBWs $\A_{\exists O. \psi}^{\le v_1}$ and $\A_\psi^{\geq v_2}$, as in the proof of Theorem~\ref{thm: FLTL os with threshold 2exp}. 
		
		Aiming to proceed Safralessly, we can also construct a UCW for $L_v$, as follows. First, note that by the monotonicity of $\comb$, for every $\zug{v_1,v_2}\in V(\psi)\times V(\psi)$ we have that $\zug{v_1,v_2}\in G_v$ iff for every $\zug{u_1,u_2}\in V(\psi)\times V(\psi) \setminus G_v$, we have that $v_1<u_1$ or $v_2>u_2$. Hence, 	
		\[
		L_v=\bigcap_{\zug{u_1,u_2}\in V(\psi)\times V(\psi)\setminus G_v}\{w\in \IOo: \sem{w,\exists O.\psi}< u_1 \mbox{ or } \sem{w,\psi}> u_2 \}, 
		\]
		and so by dualization we have 
		\[
		\IOo \setminus L_v=\bigcup_{\zug{u_1,u_2}\in V(\psi)\times V(\psi)\setminus G_v}\{w\in \IOo: \sem{w,\exists O.\psi} \geq u_1 \mbox{ and } \sem{w,\psi}\leq u_2 \}.
		\]
		Hence, we can obtain a UCW for $L_v$ by dualizing an NBW that is the union of NBWs $\A_{u_1,u_2}$, for all $\zug{u_1,u_2}\in V(\psi)\times V(\psi)\setminus G_v$, each of which is the product of NBWs $\A_{\exists O. \psi}^{\ge u_1}$ and $\A_\psi^{\leq u_2}$. 
		
		Observe that in all cases, the size of the NBW is $2^{O(|\psi|)}$. Indeed, there are at most $2^{2|\psi|}$ pairs in the union, and, by Theorem~\ref{from abk}, the size of the NBW for each pair is $2^{O(|\psi|)}$.
		
		The lower bound follows from the 2EXPTIME-hardness of LTL \ges-realizability.  
		\qed
	\end{proof}
	
	By Theorem~\ref{from abk}, the number of possible satisfaction values for $\psi$ is at most $2^{|\psi|}$. Thus, the number of possible values for $\comb(A,G)$, where $A$ and $G$ are satisfaction values of $\psi$, is at most $2^{2|\psi|}$. Using binary search over the image of $\comb$, we can use Theorem~\ref{thm: quant opt synth decision} to obtain the following.
	
	\begin{corollary}
		\label{cor: quant opt synth}
		The AG \ges-synthesis problem can be solved in doubly-exponential time.
	\end{corollary}

	\begin{remark}
		\label{rmk: comb captures guarantees}
		{\bf [\ges-synthesis as a Special Case of Assume Guarantee \ges-Synthesis]} The two approaches taken in Section~\ref{sec: FLTL optimal synt} can be captured by an appropriate $\comb$ function. Indeed, for \ges-synthesis with a threshold, we can use the function $\comb$ with $\comb(A,G)=1$ if $A\ge v\to G\ge v$, and $\comb(A,G)=0$ otherwise. For \ges-synthesis (without a threshold), we can use the function $\comb$ with $\comb(A,G)=1$ if $A=G$, and $\comb(A,G)=0$ otherwise (recall that $A\ge G$ by definition). 
		However, the solution described  in Section~\ref{sec: FLTL optimal synt} is simpler than the one described here for the general case.\hfill \qed
	\end{remark}
	
	\subsection{\FLTL  \ges-synthesis in Stochastic Environments}
	\label{sec: optimal synt stochastic}
	
	The setting of \FLTL \ges-synthesis studied in Sections~\ref{fltl os} and~\ref{fltl ag os} takes the different satisfaction values into an account, but is binary, in the sense that a specification is either  (possibly AG) \ges-realizable, or is not. In particular, in case the specification is not \ges-realizable, synthesis algorithms only return ``no".  In this section we add a quantitative measure also to the underlying realizability question. We do so by assuming a stochastic environment, with a known distribution on the inputs sequences, and analyzing the expected performance of the system.
	
	For completeness, we remind the reader of some basics of probability theory. For a comprehensive reference see e.g.,~\cite{She02}.
	Let $\Sigma$ be a finite alphabet, and let $\nu$ be some {\em probability distribution\/} over $\Sigma^\omega$. 
	%Thus, $\nu:\Sigma^\omega \rightarrow [0,1]$ is such that $\sum_{w \in \Sigma^\omega} \nu(w) = 1$. 
	For example, in the uniform distribution over $(2^I)^\omega$, the probability space is induced by sampling each letter with probability $2^{-|I|}$, corresponding to settings in which each signal in $I$ always holds in probability $\frac{1}{2}$. 
	We assume $\nu$ is given by a finite Markov Decision Process (MDP). That is, $\nu$ is induced by the distribution of each letter $i\in 2^I$ at each time step, determined by a finite stochastic control process that takes into account also the outputs generated by the system (see~\cite{AK16} for the precise model). A {\em random variable\/} is then a function $X:\Sigma^\omega\to \RR$. When $X$ has a finite image $V$, which is the case in our setting, its \emph{expected value} is $\EE[X]=\sum_{v\in V}v\cdot \Pr(X^{-1}(v))$. Intuitively, $\EE[X]$ is the ``average'' value that $X$ attains.
	Next, consider an {\em event\/} $E\subseteq \Sigma^\omega$. The \emph{conditional expectation of $X$ with respect to $E$} is $\EE[X|E]=\frac{\EE[{\mathbbm 1}_E X]}{\Pr(E)}$, where ${\mathbbm 1}_E X$ is the random variable that assigns $X(w)$ to $w\in E$ and $0$ to $w\not \in E$. Intuitively, $\EE[X|E]$ is the average value that $X$ attains when restricting to words in $E$, and normalizing according to the probability of $E$ itself.
	
	%and let $\tup{\Sigma^\omega,{\cal F},\Pr}$ be a probability space, where $\Sigma^\omega$ is the domain, ${\cal F}\subseteq 2^{\Sigma^\omega}$ is a $\sigma$-algebra over $\Sigma^\omega$ whose elements are called \emph{events}, and $\Pr$ is a probability function. For the purpose of this section, the sets in ${\cal F}$ are generated by cylinder sets of the form $C_u\{u\cdot v: v\in \Sigma^\omega \}$ for every $u\in \Sigma^*$, and we have $\Pr(C_u)=2^{-|u|}$.
	
	%A \emph{real-valued random variable} is a function $X:\Sigma^\omega\to \RR$ whose the preimages are measurable events (see~\cite{She02} for a precise definition). In our setting, the range of $X$ is a finite set $V\subseteq \RR$. The \emph{expectation} of $X$ is then $\EE[X]=\sum_{v\in V}v\cdot \Pr(X^{-1}(v))$.  
	%Consider an event $E\in {\cal F}$ with $\Pr(E)>0$. We define the \emph{conditional expectation of $X$ with respect to $E$} by $\EE[X|E]=\frac{\EE[{\mathbbm 1}_E X]}{\Pr(E)}$, where ${\mathbbm 1}_E X$ is the random variable that assigns $X(w)$ for $w\in E$ and $0$ otherwise.

	We continue and review the {\em high-quality synthesis problem} \cite{AK16}, where 
	%optimality 
	the \ges variant
	is not considered. There, the environment is assumed to be stochastic and we care for the expected satisfaction value of an \FLTL specification in the computations of a transducer $\T$, assuming some given distribution on the inputs sequences. Formally, let $X_{\T,\psi}: (2^I)^\omega \rightarrow \RR$ be a random variable that assigns each sequence $x \in (2^I)^\omega$ of input signals with $\sem{\T(x),\psi}$. Then, when the sequences in $(2^I)^\omega$ are sampled according to a given distribution $\nu$ of $(2^I)^\omega$, we define $\sem{\T,\psi}^\nu=\EE[X_{\T,\psi}]$. 
	Since $\nu$ is fixed, we omit it from the notation and use $\sem{\T,\psi}$ in the following.
	% As has been the case in \cite{AK16}, it is not difficult to generalize our results to more general probability distributions given by Markov Decision Processes (MDPs).  
	
	\begin{remark}{\rm 
			{\bf [Relating LTL \ges-synthesis with \FLTL Synthesis in Stochastic Environments]}
			Given an LTL formula $\psi$, we can view it as an \FLTL formula with possible satisfaction values $\{0,1\}$, apply to it high-quality synthesis {\em a-la} \cite{AK16}, and find a transducer $\T$ that maximizes $\EE[X_{\T,\psi}]$. An interesting observation is that if $\T$ \ges-realizes $\psi$, then it also maximizes $\EE[X_{\T,\psi}]$. Indeed, all input sequences that can contribute to the expected satisfaction value, do so. \hfill \qed}
	\end{remark}
	
	We introduce and study two measures for high-quality synthesis in a stochastic environment. In the first, termed \emph{expected \ges-synthesis}, all input sequences are sampled, yet the satisfaction value in each input sequence takes its hopefulness level into account. In the second, termed \emph{conditional expected \ges-synthesis}, only hopeful input sequences are sampled.
	
	We start with expected \ges-synthesis. There, instead of associating each sequence $x\in (2^I)^\omega$ with $\sem{x \otimes \T(x),\psi}$, we associate it with 
	$X^{\comb}_{\T,\psi}=\comb(\sem{x,\exists O. \psi}, \sem{x \otimes\T(x),\psi}\}$, where $\comb$ is as described in Section~\ref{subsec: assume guarantee}, thus capturing the assume-guarantee semantics of quantitative \ges-synthesis. Then, we define $\sem{\T,\psi}^{\comb}=\EE[X^\comb_{\T,\psi}]$. For example, taking $\comb$ as implication, we have $X^{\comb}_{\T,\psi}=\max\{\sem{x \otimes\T(x),\psi}, \sem{x,\forall O. \neg \psi}\}$, capturing the semantics of $(\exists O. \psi) \rightarrow \psi$.
	
	Then, in conditional expected \ges-synthesis, we consider $\exists O. \psi$ as an environment assumption, and factor it in using conditional expectation, parameterized by a threshold $v\in [0,1]$. Formally, let $\exists O.\psi \ge v$ denote the event $\{x\in (2^I)^\omega: \sem{x,\exists O.\psi}\ge v\}$. Then, we define $\sem{\T,\psi}^{\condvar(v)}=\EE[X_{\T,\psi}| \exists O. \psi\ge v]$, assuming the event $\exists O.\psi\ge v$ has a strictly positive probability.
	
	In \cite{AK16}, it is shown that the high-quality synthesis problem can be solved in doubly-exponential time, also in the presence of environment assumptions. In the solution, the first step is the translation of the involved formulas to DPWs. In order to extract from \cite{AK16} the results relevant to us, we describe them by means of {\em discrete quantitative specifications}, defined as follows. A discrete quantitative specification $\Psi$ over $I \cup O$ is given by means of a sequence $\A_1,\ldots,\A_n$ of DPWs, with $(2^{I\cup O})^\omega=L(\A_1)\supseteq L(\A_2)\supseteq\ldots \supseteq L(\A_n)$, and sequence $0 \leq v_1<\ldots <v_n \leq 1$ of values. For every $w\in (2^{I\cup O})^\omega$, the satisfaction value of $w$ in $\Psi$, denoted $\sem{w,\Psi}$, is $\max\{v_i : w \in L(\A_i)\}$. We refer to $n$ as the depth of $\Psi$.
	
	\begin{theorem}[\cite{AK16}]
		\label{thm: AK16 results}
		Consider a discrete quantitative specification $\Psi$ over $I\cup O$. Let $n$ be its depth and $m$ be the size of the largest DPW in $\Psi$. 
		For a transducer $\T$, let $X_{\T}$ be a random variable that assigns a word $x\in (2^I)^\omega$ with $\sem{x \otimes \T(x),\Psi}$.
		%For a transducer $\T$, let $X_{\T}$ be a random variable that assigns a word $x\in (2^I)^\omega$ we its value as prescribed by the specification. Then we have the following:
		\begin{compactenum}
			\item We can synthesize a transducer $\T$ that maximizes $\EE[X_{\T}]$ in time $m^n$. 
			\item Let $\B$ be a DPW over $(2^I)^\omega$ such that $\Pr(L(\B))>0$. Then, we can synthesize a transducer $\T$ that maximizes $\EE[X_{\T}|\B]$ in time $m^n\cdot k$, where $k$ is the size of $\B$.
		\end{compactenum}
	\end{theorem}
	
	We can now state the main results of this section.
	\begin{theorem}
		Consider an \FLTL formula $\psi$.
		\begin{compactenum}
			\item
			Given a function $\comb$, we can find in doubly-exponential time a transducer that maximizes $\sem{\T,\psi}^{\comb}$.
			\item
			Given a threshold $v \in [0,1]$, we can find in doubly-exponential time a transducer that maximizes $\sem{\T,\psi}^{\condvar(v)}$.
		\end{compactenum}
	\end{theorem}
	\begin{proof}
		Let $v_1<v_2<\ldots<v_n$ be the possible satisfaction values of $\psi$ (and hence also of $\exists O. \psi$ and of $\forall O.\psi$). By Theorem~\ref{from abk}, we have that $n\le 2^{|\psi|}$. For each $v_i$, we can construct a DPW $\D_{\comb(\exists O. \psi,\psi)}^{\ge v_i}$ as in Theorem~\ref{thm: quant opt synth decision}.
		It is not hard to see that the discrete quantitative specification given by the DPWs $\D_{\comb(\exists O. \psi,\psi)}^{\ge v_i}$ and the values $v_i$, for $1 \leq i \leq n$, is qual to the specification $\comb(\exists O.\psi, \psi)$. 
		Thus, by Theorem~\ref{thm: AK16 results} (1), we can find a transducer that maximizes $\EE[X_\T]$ in time $(2^{2^{O(|\psi|)}})^{2^{|\psi|}}=2^{2^{O(|\psi|)}}$.
		
		Next, given $v \in [0,1]$, we can check whether $\Pr(\exists O.\psi>v)>0$, for example by converting a DPW $\D^{\ge v}_{\exists O. \psi}$ to an MDP, and reasoning about its Ergodic-components. Then, by Theorem~\ref{thm: AK16 results} (2), we can find a transducer that maximizes $\EE[X_{\T}| \exists O\psi>v]$, in time $(2^{2^{O(|\psi|)}})^{2^{|\psi|}}\cdot 2^{2^{O(\psi)}}=2^{2^{O(|\psi|)}}$.	
		\qed
	\end{proof}
	
	\begin{corollary}
		\label{cor: expect opt synth}
		The (possibly conditional) expected \ges-synthesis problem for \FLTL can be solved in doubly-exponential time.
	\end{corollary}

	\subsection{Guarantees in High-Quality \ges-Synthesis}
	As in the Boolean setting, also in the high-quality one we would like to add to a \ges-realizing transducer guarantees and indications about the satisfaction level. As we detail below, the quantitative setting offers many possible ways to do so.
	
	\subsubsection{High-Quality \ges-Synthesis with Guarantees}
	We consider specifications of the form $\psi=\psis\wedge \psiw$, where essentially, we seek a transducer that realizes $\psis$ and (possibly AG) \ges-realizes $\psiw$. Maximizing the realization value of $\psis$ may conflict with maximizing the \ges-realization value of $\psiw$, and there are different ways to trade-off the two goals. Technically, in the decision-problem variant, we are given two thresholds $v_1,v_2\in [0,1]$, and we seek a transducer $\T$ that realizes $\psis$ with value at least $v_1$, and \ges-realizes $\psiw$ with value at least $v_2$. Then, one may start, for example, by maximizing the value $v_1$, and then find the maximal value $v_2$ that may be achieved simultaneously. Alternatively, one may prefer to maximize $v_2$, or some other combination of $v_1$ and $v_2$. Also, it is possible to decompose $\psi$ further, to several strong and weak components, each with its desired threshold. 
	
	The solutions in the different settings all involve a construction of a UCW $\A^{\ge v_1}_{\psis}$, and its product with the automata constructed in the solutions for the different \ges-synthesis variants. We thus have the following. We note that when the solution for $\psiw$ is Safraless, we can use a UCW for $\psis$ to maintain a Safraless construction.
	
	\begin{theorem}
		The problem of \FLTL high-quality \ges-synthesis with a guarantee can be solved in doubly-exponential time.
	\end{theorem}

	\subsubsection{Flags by a High-Quality \ges-Realizing Transducer}
	In the quantitative setting, we parameterized the flags raised by the \ges-realizing transducer by values in $[0,1]$, indicating the announced satisfaction level. 
	Thus, rather than talking about prefixes being green, red, or blue, we talk about them being $v$-green, $v$-red, and $v$-blue, for $v \in [0,1]$, which essentially means that a satisfaction value of at least $v$ is guarantees (in green and blue flags) or is impossible (in red ones). We can think of those as ``degrees'' of green, red, and blue. Below we formalize this intuition and argue that even an augmentation of a transducer that \ges-realizes $\psi$ by flags for all values in $V(\psi)$ leaves the problem in doubly-exponential time. 
	
	A {\em quantitative language\/} over $2^{I \cup O}$ is $L:\IOo\to [0,1]$. In particular, an \FLTL formula $\psi$ defines the quantitative language $L_\psi$, where for all $w \in \IOo$, we have $L_\psi(w)=\sem{w,\psi}$.
	For a quantitative language $L$ and a word $w\in (2^{I\cup O})^*$, we define $L^w$ as the quantitative language where for all $w' \in \IOo$, we have $L^w(w')=L(w\cdot w')$. 
	For a value $v \in [0,1]$, a word $w \in (2^{I \cup O})^*$ is {\em $v$-green for\/} $L$ if $L^{w}$ is $v$-realizable. That is, there is a transducer $\T$ such that $\sem{T,L^w} \geq v$. 
	A word $x \in (2^I)^*$ is {\em $v$-green for\/} $L$ if there is $y \in (2^O)^*$ such that $x \otimes y$ is $v$-green for $L$.
	Thus, when the environment generates $x$, the system can respond in a way that would guarantee $v$-realizability. 
	Finally, we say that $L$ is \emph{green realizable} if there is a strategy $f:(2^I)^+ \rightarrow 2^O$ that for every threshold $v$ and for every input $x\in (2^I)^+$ that is $v$-green for $L$, we have that $x\otimes f(x)$ is $v$-green for $L$. It is not hard to see that Theorem~\ref{thm: optimal real implies flags} carries over to the quantitative setting:\footnote{Recall that quantitative optimal synthesis can be parameterized by a threshold $v \in [0,1]$. Likewise, we can parameterize green realizability by a threshold. Then, optimal realizability with threshold $v$ implies green realizability with threshold $v$.}
	\begin{theorem}
		\label{thm: quantitative optimal real implies flags}
		Quantitative optimal realizability is strictly stronger than quantitative green realizability. In particular, if a transducer $\T$ optimally realizes an \FLTL formula $\psi$, then $\T$ also green realizes $\psi$. 
	\end{theorem}
	
	For the detection of $v$-green prefixes, we parameterize by values also the notion of universal satisfiability, and say that $L$ is $v$-universally-satisfiable if all input sequences are $v$-hopeful. Then, a word $w\in (2^{I\cup O})^*$ is \emph{light $v$-green for} $L$ if $L^w$ is $v$-universally-satisfiable, and $x\in (2^{I})^*$ is \emph{light $v$-green for $L$} if there exists $y\in (2^{O})^*$ such that $x\otimes y$ is light $v$-green for $L$. The following quantitative analogue of Lemma~\ref{lem: light green DFAs} can then be proved \emph{mutatis-mutandis}: 
	Given an \FLTL formula $\psi$ over $I\cup O$ and a threshold $v\in [0,1]$, we can construct a DFA $\S^v_\psi$ of size $2^{2^{|\psi|}}$ such that $L(\S^v_\psi)=\{x\in (2^I)^*: x \mbox{ is light $v$-green for }L_\psi\}$. Now, for languages that are  \ges-realizable with threshold $v$, we have that $v$-green and light $v$-green coincide. Accordingly, we can use monitors $\S^v_\psi$ for all desired values $v \in [0,1]$. Note that if a transducer \ges-realizes $\psi$ with threshold $v$, then we are interested in a monitor for $v$ (or lower values), and if a transducer \ges-realizes $\psi$ without a threshold, then all monitors are of potential interest. Thus, the approximation of $v$-green by light $v$-green is sound.
	Finally, recall that $\psi$ has at most $2^{|\psi|}$ possible satisfaction values, and so even an execution of all monitors is still only doubly-exponential.
	In fact, the monitors for the different thresholds share the same state space and differ only in the definition of accepting states.
	
	We continue to quantitative red and blue prefixes and flags.  
	A word $w \in (2^{I \cup O})^*$ is {\em $v$-red for\/} $L$ if $L^{w}(w')\leq v$ for every $w'\in \IOo$.
	A word $x \in (2^I)^*$ is {\em $v$-red for\/} $L$ if for all $y \in (2^O)^*$, we have that $x \otimes y$ is $v$-red for $L$.
	Thus, when the environment generates $x$, then no matter how the system responds, $L$ is satisfied with value at most $v$.
	Similarly, a word $w \in (2^{I \cup O})^*$  is \emph{$v$-blue for} $L$ if $L^{w}(w')\ge v$ for every $w'\in \IOo$. A word $x \in (2^I)^*$ is \emph{$v$-blue for} $L$ if there is $y\in (2^O)^*$ such that $x\otimes y$ is $v$-blue for $L$. Then, when the environment generates $x$, the system can respond in a way that guarantees satisfaction value $v$ no matter how the interaction continues. The study of quantitative red and blue realizability and flags proceeds with no surprises as in the Boolean setting.

	\section{Discussion}
	We introduced and solved several variants of \ges-synthesis. Our complexity results are tight and show that \ges-synthesis is not more complex than traditional synthesis. In practice, however, traditional synthesis algorithms do not scale well, and much research is devoted for the development of methods and heuristics for coping with the implementation challenges of synthesis.  A natural future research direction is to extend these heuristics and methods for \ges-synthesis.  We mention here two specific examples. 
	
Efficient synthesis algorithms have been developed for fragments of LTL  \cite{ALM03b}. Most notable is the {\em GR(1) fragment}~\cite{PPS06}, which supports assume-guarantee reasoning, and for
which synthesis has an efficient symbolic solution. Adding existential quantification to GR(1) specifications, which is how we handled LTL \ges-synthesis, is not handled by its known algorithms, and is an interesting challenge.
The success of SAT-based model-checking have led to the development of SAT-based synthesis algorithms \cite{BEKKL14}, where the synthesis problem is reduced to satisfiability of a QBF formula. The fact the setting already includes quantifiers suggests it can be extended to \ges-synthesis. A related effort is {\em bounded synthesis\/} algorithms \cite{SF07,KLVY11}, where the synthesized systems are assumed to be of a bounded size and can be represented synbolically \cite{Ehl10}.

	\stam{
	 tight complexity bounds on \ges-synthesis, and shows that tackling it can be done using similar methods to traditional synthesis. In practice, however, traditional synthesis algorithms do not scale well, and so other methods and further assumptions are employed. A natural research direction is to adapt \ges-synthesis to these methods. 
	Of particular interest is synthesis GR(1) specification~\cite{PPS06}, where the specification 
	%takes on the form $A\to G$ where $A$ and $G$ are LTL formulas of 
	has a special structure, and which admits a polynomial time algorithm. Unfortunately, the introduction of quantifiers as per Theorem~\ref{thm: ltl os 2exp} to GR(1) specifications is not handled by its known algorithms. It is thus interesting to study whether a polynomial time algorithm can be recovered.
	}
	\small
	\bibliography{../ok}

\begin{thebibliography}{10}

\bibitem{ABK16}
S.~Almagor, U.~Boker, and O.~Kupferman.
\newblock Formalizing and reasoning about quality.
\newblock {\em Journal of the ACM}, 63(3), 2016.

\bibitem{AK16}
S.~Almagor and O.~Kupferman.
\newblock High-quality synthesis against stochastic environments.
\newblock In {\em Proc. 25th Annual Conf. of the European Association for
  Computer Science Logic}, volume~62 of {\em LIPIcs}, pages 28:1--28:17, 2016.

\bibitem{AKRV17}
S.~Almagor, O.~Kupferman, J.O. Ringert, and Y.~Velner.
\newblock Quantitative assume guarantee synthesis.
\newblock In {\em Proc. 29th Int. Conf. on Computer Aided Verification}, volume
  10427 of {\em Lecture Notes in Computer Science}, pages 353--374. Springer,
  2017.

\bibitem{BCHJ09}
R.~Bloem, K.~Chatterjee, T.~Henzinger, and B.~Jobstmann.
\newblock Better quality in synthesis through quantitative objectives.
\newblock In {\em Proc. 21st Int. Conf. on Computer Aided Verification}, volume
  5643 of {\em Lecture Notes in Computer Science}, pages 140--156. Springer,
  2009.

\bibitem{BCJ18}
R.~Bloem, K.~Chatterjee, and B.~Jobstmann.
\newblock Graph games and reactive synthesis.
\newblock In {\em Handbook of Model Checking.}, pages 921--962. Springer, 2018.

\bibitem{BEKKL14}
R.~Bloem, U.~Egly, P.~Klampfl, R.~K{\"{o}}nighofer, and F.~Lonsing.
\newblock Sat-based methods for circuit synthesis.
\newblock In {\em Proc. 14th Int. Conf. on Formal Methods in Computer-Aided
  Design}, pages 31--34. {IEEE}, 2014.

\bibitem{BE98}
A.~Borodin and R.~El-Yaniv.
\newblock {\em Online Computation and Competitive Analysis}.
\newblock Cambridge University Press, 1998.

\bibitem{CHJ08}
K.~Chatterjee, T.~Henzinger, and B.~Jobstmann.
\newblock Environment assumptions for synthesis.
\newblock In {\em Proc. 19th Int. Conf. on Concurrency Theory}, volume 5201 of
  {\em Lecture Notes in Computer Science}, pages 147--161. Springer, 2008.

\bibitem{Chu63}
A.~Church.
\newblock Logic, arithmetics, and automata.
\newblock In {\em Proc. Int. Congress of Mathematicians, 1962}, pages 23--35.
  Institut Mittag-Leffler, 1963.

\bibitem{Ehl10}
R.~Ehlers.
\newblock Symbolic bounded synthesis.
\newblock In {\em Proc. 22nd Int. Conf. on Computer Aided Verification}, volume
  6174 of {\em Lecture Notes in Computer Science}, pages 365--379. Springer,
  2010.

\bibitem{FKL10}
D.~Fisman, O.~Kupferman, and Y.~Lustig.
\newblock Rational synthesis.
\newblock In {\em Proc.\ 16th Int. Conf. on Tools and Algorithms for the
  Construction and Analysis of Systems}, volume 6015 of {\em Lecture Notes in
  Computer Science}, pages 190--204. Springer, 2010.

\bibitem{Kup18}
O.~Kupferman.
\newblock Automata theory and model checking.
\newblock In {\em Handbook of Model Checking}, pages 107--151. Springer, 2018.

\bibitem{KLVY11}
O.~Kupferman, Y.~Lustig, M.Y. Vardi, and M.~Yannakakis.
\newblock Temporal synthesis for bounded systems and environments.
\newblock In {\em Proc. 28th Symp. on Theoretical Aspects of Computer Science},
  pages 615--626, 2011.

\bibitem{KPV16}
O.~Kupferman, G.~Perelli, and M.Y. Vardi.
\newblock Synthesis with rational environments.
\newblock {\em Annals of Mathematics and Artificial Intelligence}, 78(1):3--20,
  2016.

\bibitem{KPV06}
O.~Kupferman, N.~Piterman, and M.Y. Vardi.
\newblock Safraless compositional synthesis.
\newblock In {\em Proc. 18th Int. Conf. on Computer Aided Verification}, volume
  4144 of {\em Lecture Notes in Computer Science}, pages 31--44. Springer,
  2006.

\bibitem{KV01d}
O.~Kupferman and M.Y. Vardi.
\newblock Model checking of safety properties.
\newblock {\em Formal Methods in System Design}, 19(3):291--314, 2001.

\bibitem{KV05c}
O.~Kupferman and M.Y. Vardi.
\newblock Safraless decision procedures.
\newblock In {\em Proc.\ 46th IEEE Symp. on Foundations of Computer Science},
  pages 531--540, 2005.

\bibitem{PPS06}
N.~Piterman, A.~Pnueli, and Y.~Saar.
\newblock Synthesis of reactive(1) designs.
\newblock In {\em Proc. 7th Int. Conf. on Verification, Model Checking, and
  Abstract Interpretation}, volume 3855 of {\em Lecture Notes in Computer
  Science}, pages 364--380. Springer, 2006.

\bibitem{Pnu81}
A.~Pnueli.
\newblock The temporal semantics of concurrent programs.
\newblock {\em Theoretical Computer Science}, 13:45--60, 1981.

\bibitem{PR89a}
A.~Pnueli and R.~Rosner.
\newblock On the synthesis of a reactive module.
\newblock In {\em Proc.\ 16th ACM Symp. on Principles of Programming
  Languages}, pages 179--190, 1989.

\bibitem{ALM03b}
P.~Madhusudan R.~Alur, S. La~Torre.
\newblock Playing games with boxes and diamonds.
\newblock In {\em Proc. 14th Int. Conf. on Concurrency Theory}, volume 2761 of
  {\em Lecture Notes in Computer Science}, pages 127--141. Springer, 2003.

\bibitem{Ros92}
R.~Rosner.
\newblock {\em Modular Synthesis of Reactive Systems}.
\newblock PhD thesis, Weizmann Institute of Science, 1992.

\bibitem{Saf88}
S.~Safra.
\newblock On the complexity of $\omega$-automata.
\newblock In {\em Proc.\ 29th IEEE Symp. on Foundations of Computer Science},
  pages 319--327, 1988.

\bibitem{SF07}
S.~Schewe and B.~Finkbeiner.
\newblock Bounded synthesis.
\newblock In {\em 5th Int. Symp. on Automated Technology for Verification and
  Analysis}, volume 4762 of {\em Lecture Notes in Computer Science}, pages
  474--488. Springer, 2007.

\bibitem{She02}
Ross Sheldon.
\newblock {\em A first course in probability}.
\newblock Pearson Education India, 2002.

\bibitem{SVW87}
A.P. Sistla, M.Y. Vardi, and P.~Wolper.
\newblock The complementation problem for {B\"uchi} automata with applications
  to temporal logic.
\newblock {\em Theoretical Computer Science}, 49:217--237, 1987.

\bibitem{VW94}
M.Y. Vardi and P.~Wolper.
\newblock Reasoning about infinite computations.
\newblock {\em Information and Computation}, 115(1):1--37, 1994.

\bibitem{Win71}
D.~W. Winnicott.
\newblock {\em Playing and Reality}.
\newblock Penguin, 1971.

\bibitem{Wol81}
P.~Wolper.
\newblock Temporal logic can be more expressive.
\newblock In {\em Proc.\ 22nd IEEE Symp. on Foundations of Computer Science},
  pages 340--348, 1981.

\end{thebibliography}
	\bibliographystyle{plain}

	\appendix
	
	\section{Counterstrategies in \ges-Synthesis}
	\label{app dual}
	
	An environment strategy is $g:(2^O)^* \rightarrow 2^I$. For an output sequence $y=o_0 \cdot o_1 \cdot o_2 \cdots \in (2^O)^\omega$, we use $g(y)$ to denote the input sequence  $g(\epsilon)\cdot g(o_0) \cdot g(o_0 \cdot o_1) \cdot g(o_0 \cdot o_1 \cdot o_2) \cdots \in (2^I)^\omega$. Then, $g(y) \otimes y \in (2^{I \cup O})^\omega$ is the {\em computation\/} of $g$ on $y$. Note that the setting is not completely dual to that of a strategy of the system, as in both cases the  
	environment initiates the interaction, and thus $g(\epsilon)$ is prepended. We say that a specification $\psi$ is $O/I$-realizable if there is an environment strategy $g:(2^O)^* \rightarrow 2^I$ such that for all output sequences $y \in (2^O)^\omega$, the computation of $g$ on $y$ satisfies $\psi$. 
	
	Determinancy of games implies that in traditional synthesis, a specification $\psi$ is not $I/O$-realizable iff $\neg \psi$ is $O/I$-realizable.
	For \ges-synthesis, simple dualization does not hold, but we do have determinancy in the following sense.
	We say that an environment strategy $g:(2^O)^* \rightarrow 2^I$ {\em \ges-realizes} $\psi$ if all $y \in (2^I)^\omega$ are hopeful with respect to $\neg \psi$ and $g$ realizes $\psi$. 
	\begin{theorem}
		\label{determinancy}
		For every specification $\psi$, we have that $\psi$ is \ges-realizable by the system iff $\neg \psi$ is not \ges realizable by the environment. 
	\end{theorem}
	
	\begin{proof}
		By determinancy of games, we have that $\psi \vee \forall O. \neg \psi$ is $I/O$-realizable iff $\neg \psi \wedge \exists O. \psi$ is not $O/I$-realizable. 
		Hence, $\psi$ is \ges realizable by the system iff $\psi \vee \forall O. \neg \psi$ is $I/O$-realizable iff $\neg \psi \wedge \exists O. \psi$ is not $O/I$-realizable iff $\neg \psi$ is not \ges realizable by the environment. \hfill \qed
	\end{proof}
	
	By Theorem~\ref{determinancy}, we can solve the synthesis problem for the dual setting in 2EXPTIME by constructing an automaton for $L=\{w: w \models \neg \psi \wedge \exists O. \psi\}$. We note that while a construction of an NBW for $L$ follows the same lines as the construction in the proof of Theorem~\ref{thm: ltl os 2exp}, the problem of generating a UCW, aiming to proceed Safralessly, is open.

	\section{Red and Blue Flags}
	\label{redblue}
	
	Consider a language $L \subseteq (2^{I \cup O})^\omega$.
	We say that a strategy $f:(2^I)^* \rightarrow 2^O$ {\em red realizes\/} $L$ if for every $x \in (2^I)^*$, if $x$ is not red for $L$, then $x \otimes f(x)$ is not red for $L$. Then, $f$ {\em blue realizes\/} $L$ if for every $x \in (2^I)^*$, if $x$ is blue for $L$, then $x \otimes f(x)$ is blue for $L$.
	
	%We say that a language $L \subseteq (2^{I \cup O})^\omega$ is {\em red realizable\/} if there is a strategy $f:(2^I)^* \rightarrow 2^O$ such that for every $x \in (2^I)^*$, if $x$ is not red for $L$, then $x \otimes f(x)$ is not red for $L$. Then, $L$ is {\em blue realizable\/} if there is a strategy $f$ such that for every $x \in (2^I)^*$, if $x$ is blue for $L$, then $x \otimes f(x)$ is blue for $L$.

	\begin{theorem}
		\ges realization implies red realization and may conflict with imply blue realization.  Neither red nor blue realization imply \ges realization.
	\end{theorem}
	
	\begin{proof}
		We first prove that every $f:(2^I)^+ \rightarrow 2^O$ that \ges realizes $\psi$ also red realizes $\psi$. Consider $x \in (2^I)^+$  that is not red for $\psi$. By definition, there is $y \in (2^O)^+$ such that $L^{x \otimes y}\neq \emptyset$. Then, there is $x' \in (2^I)^\omega$ and $y' \in (2^O)^\omega$ such that $x' \otimes y'$ in $L^{x \otimes y}$. Hence, $x \cdot x'$ is hopeful for $\psi$. Therefore, as $f$ \ges realizes $\psi$, we have that $(x \cdot x') \otimes f(x \cdot x') \models \psi$. Thus, $x \otimes f(x)$ is not red, and so $f$ red realizes $\psi$. 
		
		We continue to an example where \ges realization conflicts with blue realization. Let $I=\{p\}$ and $O=\{q\}$, and 
		consider the specification $\psi= (q \wedge \Next(p \vee \psi_1)) \vee (\neg q \wedge \Next \psi_2)$, where $\psi_1$ is not \ges realizable and $\psi_2$ is \ges realizable. The input sequence $\emptyset\cdot \{p\}$ is blue, as $\emptyset\cdot \{p\} \otimes \{q\}\cdot \emptyset=\{q\}\cdot\{p\}$ is blue for $\psi$. 
		The specification $\psi$ is \ges realizable: a strategy $f:(2^I)^+ \rightarrow 2^O$ that \ges realizes $\psi$ has $f(i_0)=\emptyset$, and then proceed to \ges realize $\psi_2$. Yet, every strategy $f$ that \ges realizes $\psi$ does not blue realizes it. Indeed, the strategy cannot have $f(i_0)=\{q\}$, as then an environment that responds (in the second input) with $\emptyset$ forces $f$ to \ges realize $\psi_1$. Hence, $f(i_0)$ must be $\emptyset$, but
		\[\emptyset\cdot \{p\} \otimes \emptyset\cdot f(\emptyset\cdot \{p\})=\emptyset\cdot (\{p\}\cup f(\emptyset\cdot \{p\})),\] which, as a word in $(2^{I \cup O})^*$, is not blue for $\psi$.
		
		Now, for the second claim, we use the same specification we used in the proof of Theorem~\ref{thm: optimal real implies flags}. Let $I=\{p\}$ and $O=\{q\}$, and 
		consider the specification $\psi= \Alw \Ev((\Next p) \wedge q) \wedge \Alw \Ev((\Next \neg p) \wedge \neg q)$. While $\psi$ is not \ges realizable, no prefix is red or blue for it, and thus all strategies red and blue realize it.
		\qed
	\end{proof}
	
	\section{Examples of $v$-Good Strategies}
	\label{apx:example helpful}
	We analyze in detail a more generalized version of Example~\ref{v-helpful example}.

	Let $I=\{p\}$ and $O=\{q\}$. Consider the \FLTL formula $\psi=(\factorU_{\lambda_1} p \vee \factorU_{\lambda_2}q)$. Checking for which values $v$ a strategy $f$ is $v$-good for $x$ with respect to $\psi$, we examine whether $\sem{x \otimes f(x), \factorU_{\lambda_1} p \vee \factorU_{\lambda_2}q} \geq v$ or $\sem{x, \forall q.\neg( \factorU_{\lambda_1} p \vee \factorU_{\lambda_2}q)} >1-v$.
	Since $\psi$ refers only to the first position in the computation, we can distinguish between four possible cases:
	\begin{itemize}
		\item
		$x_0=\emptyset$ and $f(x_0)=\emptyset$. 
		Then, 
		$\sem{x \otimes f(x), \factorU_{\lambda_1} p \vee \factorU_{\lambda_2}q} = 0$, $\sem{x,\exists q.\factorU_{\lambda_1} p \vee \factorU_{\lambda_2}q}=\max\{0,\lambda_2\}=\lambda_2$, and 
		$\sem{x, \forall q. \neg(\factorU_{\lambda_1} p \vee \factorU_{\lambda_2}q)}=\min\{1,1-\lambda_2\}=1-\lambda_2$.
		Hence, $f$ is $v$-good for $x$ with respect to $\psi$ if $v = 0$ or $v > \lambda_2$, thus $v \in \{0\} \cup (\lambda_2,1)$
		\item
		$x_0=\emptyset$ and $f(x_0)=\{q\}$. 
		Then, 
		$\sem{x \otimes f(x), \factorU_{\lambda_1} p \vee \factorU_{\lambda_2}q} = \lambda_2$, $\sem{x,\exists q.\factorU_{\lambda_1} p \vee \factorU_{\lambda_2}q}=\lambda_2$, and $\sem{x, \forall q. \neg(\factorU_{\lambda_1} p \vee \factorU_{\lambda_2}q)}=1-\lambda_2$. 
		Hence, $f$ is $v$-good for $x$ with respect to $\psi$ if $v \leq \lambda_2$ or $1-v < 1-\lambda_2$, thus for all $v \in [0,1]$.
		\item
		$x_0=\{p\}$ and $f(x_0)=\emptyset$. 
		Then, 
		$\sem{x \otimes f(x), \factorU_{\lambda_1} p \vee \factorU_{\lambda_2}q} = \lambda_1$, $\sem{x,\exists q.\factorU_{\lambda_1} p \vee \factorU_{\lambda_2}q}=\max\{\lambda_1,\lambda_2\}$, and $\sem{x, \forall q. \neg(\factorU_{\lambda_1} p \vee \factorU_{\lambda_2}q)}=\min\{1-\lambda_1,1-\lambda_2\}$. Hence, $f$ is 
		$v$-good for $x$ with respect to $\psi$ if $v \leq \lambda_1$ or $1-v < \min\{1-\lambda_1,1-\lambda_2\}$, thus $v \in [0,\lambda_1] \cup (\max\{\lambda_1,\lambda_2\},1]$. When $\lambda_1 < \lambda_2$, this means that $v \in [0,\lambda_1] \cup (\lambda_2,1]$. When $\lambda_1 \geq \lambda_2$, this means that $v \in [0,1]$. 
		\item
		$x_0=\{p\}$ and $f(x_0)=\{q\}$. 
		Then, 
		$\sem{x \otimes f(x), \factorU_{\lambda_1} p \vee \factorU_{\lambda_2}q} = \max\{\lambda_1,\lambda_2\}$, $\sem{x,\exists q.\factorU_{\lambda_1} p \vee \factorU_{\lambda_2}q}=\max\{\lambda_1,\lambda_2\}$, and $\sem{x, \forall q. \neg(\factorU_{\lambda_1} p \vee \factorU_{\lambda_2}q)}=\min\{1-\lambda_1,1-\lambda_2\}$. Hence, $f$ is 
		$v$-good for $x$ with respect to $\psi$ if $v \leq \max\{\lambda_1,\lambda_2\}$ or $1-v < \min\{1-\lambda_1,1-\lambda_2\}$, thus for all $v \in [0,1]$.
	\end{itemize}

\end{document}